\documentclass[12pt,letterpaper]{article} 
\input epsf
\usepackage{fontenc}
\usepackage{epsfig}
\usepackage{rotating}
\usepackage{graphicx}
\usepackage{setspace}
\usepackage{amsmath}
\usepackage{amssymb}
\usepackage{multirow}
\epsfverbosetrue
\def\abstract{\if@twocolumn
\section*{Abstract}
\else \normalsize 
\begin{center}
{\bf Summary\vspace{-.5em}\vspace{0pt}} 
\end{center}
\quotation 
\fi}
\def\endabstract{\if@twocolumn\else\endquotation\fi}

\makeatletter
\@addtoreset{equation}{section}
\makeatother

\newtheorem{remark}{Remark}[section]

\newcommand{\myappendix}[1]{
	\setcounter{section}{1}
        \renewcommand{\thesection}{A\arabic{section}}}

\begin{document}
\onehalfspacing
\newpage
\pagestyle{plain}
%%%%%%%%%%%%%%%%%%%%%%%%%%%%%%%%%%%%%%%%%%%%%%%
\title{Bayesian Approaches to Copula Modelling}
%%%%%%%%%%%%%%%%%%%%%%%%%%%%%%%%%%%%%%%%%%%%%%

\author{ Michael Stanley Smith\\
{\em (Melbourne Business School, University of Melbourne)}}

\maketitle
\vspace{2in}
\begin{center}
(To appear in:
Paul Damien, Petros Dellaportas, Nicholas Polson, and David Stephens (Eds).
{\em Hierarchical Models and MCMC: A Tribute to Adrian Smith}.)
\newpage

\vspace{1in}
{\Large {\bf Abstract}}
\end{center}
\noindent
Copula models have become
one of the most widely used tools in the applied modelling
of multivariate data.
Similarly, Bayesian methods are increasingly used
to obtain efficient likelihood-based 
inference. However, to date,
there has been only limited use of Bayesian approaches in the formulation
and estimation of copula models. This article aims to address this shortcoming in two ways.
First,
to introduce copula models and
aspects of copula theory that are especially relevant for a Bayesian analysis.
Second, to outline Bayesian approaches to formulating and estimating
copula models, and their advantages over alternative methods. 
Copulas covered include Archimedean, copulas constructed
by inversion, and vine copulas; along with their interpretation as
transformations.
A number of parameterisations of a correlation
matrix of a
Gaussian copula are considered, along with
hierarchical priors that allow for 
Bayesian selection and model averaging for each parameterisation.
Markov chain Monte Carlo sampling schemes 
for fitting Gaussian and D-vine copulas,
with and without selection,
are given in detail. The relationship between the prior for the parameters
of a D-vine, and the prior for a correlation matrix of a Gaussian copula,
is discussed.
Last, it is shown how to compute Bayesian inference when the data are discrete-valued
using data augmentation. This approach
generalises popular Bayesian methods for the estimation of
models for multivariate binary and other ordinal data to more general copula models.
Bayesian data augmentation has
substantial advantages over other methods of estimation for
this class of models.
\newpage

\section{Introduction}

Copula models are now used widely in the empirical analysis of multivariate data.
For example,
major areas of application include survival analysis, where
much early work occurred (Clayton 1978; Oakes 1989), actuarial science (Frees
and Valdez~1998), finance (Li~2000; 
Cherubini, Luciano and Vecchiato~2004; McNeil, Frey and Embrechts~2005),
marketing (Danaher and Smith~2011), transport studies
(Bhat and Eluru~2009; Smith and Kauermann~2011), medical statistics 
(Lambert and Vandenhende~2002; Nikoloulopoulos and Karlis~2008) and 
econometrics (Smith~2003; Cameron et al.~2004; Patton~2006).
Copula models are popular because they are flexible tools
for the modelling of complex relationships between variables in a simple manner.
They allow for the marginal distributions of data
to be modelled separately in an initial step, and then
dependence between variables
is captured using a copula function.

However, the
development of estimation and statistical inferential methodology for
copula models has been limited. 
Most research has either been focused on the development and
properties of copula
functions (see Joe~1997 and Nelsen~2006 for excellent overviews), 
or their use in solving applied problems. 
Less attention has been given to the question of how to estimate the 
increasing variety of copula models in an effective manner.
To date, the most popular estimation methods are full or two-stage 
maximum likelihood estimation (Joe~2005) and method of moments style
estimators 
in low dimensions (Genest and Rivest~1993). There has
been only limited work on developing Bayesian approaches to formulate
and estimate copula models. This is surprising, given that 
Bayesian methods have proven
successful in both formulating and estimating
multivariate models elsewhere.
The aim of this article is two-fold:
(i)~to introduce contemporary copula modelling to Bayesian statisticians, and
(ii)~to outline the advantages of Bayesian inference when applied to 
copula models. Therefore, there are two intended audiences:
(i)~Bayesians who are unfamiliar with the advances and features 
of copula models, 
and (ii)~users of copula 
models who are unfamiliar with the advantages and features of modern
Bayesian inferential methods.
 
Previous Bayesian work on copula modelling includes that of
Huard, \'{E}vin and Favre~(2006), who suggest
a method
to select between different bivariate copulas, and
Silva and Lopes~(2008) who use Markov chain Monte Carlo (MCMC) methods to estimate low dimensional 
parametric copula functions.
Pitt, Chan and Kohn (2006), Hoff~(2007) and Danaher and Smith~(2011) 
estimate Gaussian copula
regression models using MCMC methods.
Note that adopting a Gaussian copula does not mean the data are normally 
distributed.
Smith, Gan and Kohn~(2010b) extend the work of Pitt, Chan and Kohn~(2006)
to copulas derived by inversion from skew t distributions constructed by hidden conditioning.
Smith et al.~(2010) and Min and Czado (2010; 2011) propose methods
to estimate so called `vine' copulas 
with continuous margins using MCMC. Pitt, Chan and Kohn~(2006) show
how Bayesian covariance selection approaches
can be used in Gaussian copulas, while 
Smith et al.~(2010) 
and Min and Czado~(2011) also show how Bayesian selection ideas
can be applied to determine whether, or not, the component `pair-copulas' of a vine
copula
are equal to the bivariate
independence copula. Smith et al.~(2010) also show that the D-vine copula provides
a natural decomposition for serial dependence. 
Ausin and Lopes~(2010) consider Bayesian estimation of multivariate time series
with copula-based
time varying cross-sectional dependence. Last, Smith and Khaled~(2011) suggest
efficient Bayesian data augmentation methodology for the estimation of copula models 
for multivariate
discrete data, or a combination of discrete and continuous data.
Their approach is for general copula functions, not just Gaussian 
copulas,
or copulas constructed by inversion.

This article is divided into three main sections. The first provides an
introduction to copula modelling. There are a number excellent in-depth introductions
to copulas and their properties; for example, see Joe~(1997) and
Nelsen~(2006). The purpose of this section is not to replicate any of these, but
to introduce aspects that are important
in Bayesian copula modelling. This includes an outline of what makes copula models 
so useful,
how copulas
models can be viewed as transformations, what are copulas constructed by inversion
and vine copulas, and why 
the D-vine copula is a natural model of serial dependence.

In the next two sections
Bayesian approaches to formulating and estimating copula models are discussed 
separately
for multivariate continuous and discrete data.
%in Sections~\ref{sec:ms:bifcm} and~\ref{sec:ms:discrete}, respectively.
This is because copula models, and associated methods, differ substantially in these two
cases. In Section~\ref{sec:ms:bifcm}
the advantages of using Bayesian inference over maximum likelihood
for case of continuous data are discussed. For the Gaussian copula,
a sampling scheme that can be used to evaluate the joint posterior distribution of the copula
and any marginal model parameters is outlined in detail. 
Different priors for the correlation matrix of the Gaussian copula are considered, 
including priors
based on a Cholseky factorisation, the partial correlations as in Pitt, Chan and Kohn~(2006),
and the conditional correlations discussed in Joe~(2005) and Daniels and Pourahmadi~(2009). 
A new Bayesian selection approach using the latter
is outlined, where the fitted copula model is a Bayesian
model average over parsimonious representations
of the dependence structure.
Bayesian
estimation and selection for D-vine copulas
is also outlined. An interesting insight is that Bayesian selection of individual
pair-copulas nests
Bayesian selection of
the conditional correlations for a Gaussian copula. Bayesian estimates of 
popular dependence
metrics from the fitted copula are also discussed, 
where parameter uncertainty can be integrated out 
using the Monte Carlo iterates from the sampling scheme.

Denuit and Lambert~(2005) and Genest and Ne\v{s}lehov\'{a}~(2007) point out that
popular method of 
moments
style estimators based on ranks
should not be used to estimate copula models for discrete data, making
likelihood-based inference more important. However, 
the likelihood function differs 
substantially from that in the continuous case, and computational issues mean
that maximum likelihood
estimation is more difficult than in the continuous case. 
An effective solution is to employ Bayesian data augmentation, as outlined for
a Gaussian copula in
Section~\ref{sec:ms:discrete}.
The priors for the correlation matrix of the Gaussian copula, and also the 
Bayesian selection framework, are unaffected by whether the data is 
discrete or continuous.
Last, it is discussed how measuring dependence in discrete data
differs from that in the continuous case.

\section{What Are Copula Models?}\label{sec:ms:wacm}
\subsection{The basic idea}
Consider initially the bivariate case with two random
variables, $Y_1$ and $Y_2$, with marginal distribution functions
$F_1(y_1)$ and $F_2(y_2)$, respectively. A copula model
is a way of constructing the joint distribution of 
$(Y_1,Y_2)$.
Sklar~(1959) shows that there always
exists a bivariate function $C:[0,1]^2 \rightarrow [0,1]$, such that
\[
F(y_1,y_2)=C(F_1(y_1),F_2(y_2))\,.
\]
The function $C$ is itself a distribution function 
with uniform margins on $[0,1]$, and is labelled the `copula function'.
It binds together the 
univariate margins $F_1$ and $F_2$ to produce
bivariate distribution $F$.

If both margins $F_1$ and $F_2$ are
continuous distribution functions, then there is a unique copula function $C$
for any given joint distribution function $F$. If either
$F_1$ or $F_2$ are discrete-valued, then $C$ is not unique. 
However, the objective of copula modelling
is not to find the copula function(s) $C$ that 
satisfy Sklar's representation, given knowledge of $F_1,F_2$ and $F$. Instead,
the objective is to
construct a joint distribution $F$ from a copula function 
$C$ and marginal models
for $F_1$ and $F_2$. In this way,
copula models can be used equally for discrete or continuous
data, or a combination of both.

It is important to notice that the
copula function $C$ does not determine the marginal
distributions of $F$, but
accounts for
dependence between
$Y_1$ and $Y_2$. For example, in the case where $Y_1$ and $Y_2$ are
independent,
the copula function is $C(u_1,u_2)=u_1u_2$, so that
$F(y_1,y_2)=F_1(y_1)F_2(y_2)$. This
copula function is called
the `independence copula'.

The copula model is easily generalised to $m$ dimensions as follows.
Let
$Y=(Y_1,\ldots,Y_m)\in {\cal S}_Y$ be a random vector with elements that have
marginal distribution functions $F_1,\ldots,F_m$, then the joint
distribution function of $Y$ is
\begin{equation}
F(y_1,\ldots,y_m)=C(F_1(y_1),\ldots,F_m(y_m))\,.
\label{eq:ms:sklar}
\end{equation}
Again, the copula
function
$C:[0,1]^m \rightarrow [0,1]$ is 
itself a distribution function for random vector $U=(U_1,\ldots,U_m)'$
with uniform margins on $[0,1]$. As before, if all elements of
$Y$ are continuous
random variables, then there is a unique copula function $C$ for any given
$F$, but this is not the case if one or more elements are discrete-valued. 
Nevertheless, Equation~(\ref{eq:ms:sklar}) can still be used to construct
a well-defined joint distribution $F$, given $F_1,\ldots,F_m$ and $C$, just
as in the bivariate case. 

\subsection{Why are copula models so useful?}
A key feature of the copula
representation of a joint distribution
is that it allows for the margins to be modelled 
separately from the dependence structure. This promotes a `bottom-up' modelling
strategy, where models are first developed one-by-one
for each univariate margin. Dependence is then introduced by 
an appropriate copula function $C$. Sklar's theorem
reassures that this is not an ad-hoc approach,
and that there should be at least
one
copula function $C$ that correctly constructs the joint distribution $F$,
as long
as the marginal models $F_1,\ldots,F_m$ are accurate. 
Compare this to a more restrictive `top-down' alternative,
where the
joint distribution function $F$ is selected first, which then determines the form of the 
marginals. For example, if $F$
is a multivariate t distribution with $\nu$ degrees
of freedom, then each
$F_j$ is restricted to be univariate $t$ with a common degrees of freedom $\nu$.

For much applied multivariate
modelling, the flexibility that the bottom-up approach
allows is compelling. The marginal models can be of the same form,
or completely different, including any of the
following:

\begin{itemize}
\item[(i)] {\em Parametric Distributions}: 
A parametric distribution $F_j(y_j;\theta_j)$,
with parameters $\theta_j$. 
For example, $F_j$ may be a $t$ distribution with location $\mu_j$, scale
$\sigma_j>0$ and degrees of freedom $\nu_j>0$,
so that $\theta_j=\{\mu_j,\sigma_j,\nu_j\}$. 
A copula model with $t$ distributions for each margin is more flexible than 
a multivariate $t$ distribution because the level of kurtosis can differ in each
dimension (Fang, Fang and Kotz~2002).
For discrete data, $F_j$ may be a negative
binomial distribution with stopping parameter $r_j>0$ and success
parameter $p_j\in(0,1)$, so that $\theta_j=\{r_j,p_j\}$. The negative binomial
is a
very popular model for count data that exhibit 
heterogeneity,
and copula models provide flexible multivariate extensions
(Lee~1999; Nikoloulopoulos and Karlis~2010; Danaher and Smith~2011).
\item[(ii)]
{\em Nonparametric Distributions}:
Approaches where each margin is modelled nonparametrically
using the empirical
distribution function (or a smoothed variant)
have long been advocated in the copula literature; for example, see
Genest, Ghoudi and Rivest~(1995), Shih and Louis~(1995) and Chen, Fan and
Tsyrennikov~(2006).
Similarly, $F_j$ can be modelled using
Bayesian nonparametric methods; see
Hjort et al.~(2010) for recent
accounts of these. Alternatively, rank likelihoods can be used for each
marginal model as outlined by Hoff~(2007). In all cases,
copula models provide simple
multivariate extensions of existing nonparametric methods.
\item[(iii)]
{\em Regression Models}:
Univariate regression models can be used for each margin, in which 
case the resulting copula model is called a `copula regression 
model' (Oakes \& Ritz~2000; Song~2000).
The regression coefficients $\beta_j$
can be pooled across margins $j=1,\ldots,m$,
so that $\beta_1=\beta_2=\ldots=\beta_m$,
in which case
the copula model is then an extension of
the multivariate regression model. If the regression coefficients
differ for each margin, then the copula model extends the `seemingly unrelated
regression' model popular in econometric analysis (Zellner~1962).
\item[(iv)]
{\em Time Series Models}:
When observations are made on a multivariate vector over time, the marginal
models can be parametric time series models, and contemporaneous dependence
captured via the copula function (Patton~2006; Chen and Fan~2006;
Ausin and Lopes~2010). 
Popular choices are GARCH
or stochastic volatility models for the margins.
As with copula regression
models, marginal
parameters can either be pooled or allowed to vary across margin.
\end{itemize}
\subsection{Copula functions and densities}
Nelsen~(2006, p.45) lists the three conditions that $C$ needs to meet
to be an admissible copula function, which are:
\begin{itemize}
\item[(i)] For every $u=(u_1,\ldots,u_m)\in [0,1]^m$,
$C(u)=0$ if at least one element $u_i=0$.
\item[(ii)] If all elements of $u$ are equal to one, except $u_i$, then 
$C(u)=u_i$.
\item[(iii)] For each $a=(a_1,\ldots,a_m),b=(b_1,\ldots,b_m)\in [0,1]^m$,
such that $a_i \leq b_i$ for all $i=1,\ldots,m$, 
\[\Delta_{a_m}^{b_m}\Delta_{a_{m-1}}^{b_{m-1}}\cdots\Delta_{a_1}^{b_1}C(v) \geq 0\,.\]
\end{itemize}
Here, $\Delta_{a_k}^{b_k}$ is a differencing notation defined as 
\begin{eqnarray*}
\lefteqn{\Delta_{a_k}^{b_k} C(u_1,\ldots,u_{k-1},v_k,u_{k+1},\ldots,u_m) = } \\
&\;\;\;\;\; &C(u_1,\ldots,u_{k-1},b_k,u_{k+1},\ldots,u_m)-
C(u_1,\ldots,u_{k-1},a_k,u_{k+1},\ldots,u_m)\,,
\end{eqnarray*}
with $v_k$ a variable of differencing, and $v=(v_1,\ldots,v_m)$. Notice that if 
$c(u)=\partial^m C(u)/\partial u_1\ldots\partial u_m$ exists, then 
property~(iii) is equivalent to
\[\int_{a_1}^{b_1}\cdots\int_{a_m}^{b_m} c(u)\mbox{d}u\geq 0\,.\]
Properties~(i) and~(iii) are satisfied if $C(u)$ is a distribution
function on $[0,1]^m$, while property~(ii) is satisfied if $C$ also has uniform
margins. The density function $c(u)$ is commonly referred to as the `copula density'.

In the vast majority of cases
parametric copula functions 
$C(u;\phi)$,
with parameters $\phi$, are used in applied analysis.
There are a large number of choices for $C$,
with Joe~(1997) and Nelsen~(2006) providing overviews of a wide range of copula functions
and their properties. Particularly popular in the bivariate case are the
family of
Archimedean copulas; see Nelsen~(2006; Chap.~4). Three of the most
popular Archimedean copulas are the Frank, Clayton and Gumbel. These are listed
in Table~\ref{tab:ACops},
along with their densities and measures of dependence.
\begin{table}[tb]
\begin{center}
\begin{tabular}{l} \hline \hline
Frank ($\phi \in (-\infty,0)\cup (0,\infty)$) \\ \hline
$C(u_1,u_2;\phi)= -\frac{1}{\phi}\log\left( 1+ \frac{(\exp(-\phi u_1)-1)(\exp(-\phi u_2)-1)}{\exp(-\phi)-1}\right)$ \\
$c(u_1,u_2;\phi)= \phi\left(\exp(\phi(1+u_1+u_2))(\exp(\phi)-1)\right)$\\
\hspace{2cm}$\times \left[\exp(\phi)-\exp(\phi(1+u_1))-\exp(\phi(1+u_2))+\exp(\phi(u_1+u_2))\right]^{-2}$\\
$\tau_{1,2}(\phi)=1+\frac{4}{\phi}(D_1(\phi)-1)$,
$\lambda_{1,2}^L(\phi)=\lambda_{1,2}^U(\phi)=0$ \\ \hline
Clayton ($\phi \in (-1,\infty) \backslash \{0\} $) \\ \hline
$C(u_1,u_2;\phi)=
\max \left\{ (u_1^{-\phi}+u_2^{-\phi}-1)^{-1/\phi},0\right\}$ \\
$c(u_1,u_2; \phi)\ = \max\left\{ (1+\phi)(u_1 u_2)^{-1-\phi}
\left(u_1^{-\phi}+u_2^{-\phi}-1\right)^{-1/\phi-2},0\right\}$ \\
%$C_{1|2}(u_1|u_2;\phi)=
%\max\left\{ u_2^{-(1+\phi)}\left(u_1^{-\phi}+u_2^{-\phi} -1\right)^{-(1+1/\phi)},0\right\}$ \\
%$C_{1|2}^{-1}(v|u_2;\phi) = 
% \left(1 - u_2^{-\phi} + \left[v u_2^{(1+\phi)} \right]^{-\phi/(\phi+1)} \right)^{-1/\phi}$ \\
$\tau_{1,2}(\phi)=\phi/(\phi+2)$, 
$\lambda_{1,2}^L(\phi)=2^{-1/\phi}$ and $\lambda_{1,2}^U(\phi)=0$ \\ \hline
Gumbel ($\phi \geq 1$) \\ \hline
$C(u_1,u_2;\phi)=\exp(-(\tilde u_1^{\phi}+\tilde u_2^{\phi})^{1/\phi})$\,,\;\;\mbox{where }$\tilde u_j = -\log(u_j)$\\
$c(u_1,u_2; \phi)\ = C(u_1,u_2;\phi)\,(u_1\,u_2)^{-1}(\tilde u_1^{\phi}+\tilde u_2^{\phi})^{-2+2/\phi}(\tilde u_1\tilde u_2)^{\phi-1}$\\
\hspace{2cm}$\times \left[ 1+(\phi-1)\left( \tilde u_1^\phi +\tilde u_2^\phi \right)^{-1/\phi} \right]$ \\
%$C_{1|2}(u_1|u_2;\phi)= C(u_1,u_2;\phi) \frac{1}{u_2}(\tilde u_2)^{\phi-1} \left[\tilde u_1^{\phi}+ \tilde u_2^{\phi}\right]^{1/\phi-1}$\\
%$C_{1|2}^{-1}$ : Obtained Numerically using Newton's Method \\
$\tau_{1,2}(\phi)=1-\phi^{-1}$,
$\lambda_{1,2}^L(\phi)=0$ and $\lambda_{1,2}^U(\phi)=2-2^{1/\phi}$ \\ \hline \hline
\end{tabular}
\end{center}
\caption{Copula functions, density functions and measures of dependence
for the Frank, Clayton and Gumbel copulas. For the Frank copula,
the function $D_1(\phi)=\frac{1}{\phi}\int_0^\phi t/(\exp(t)-1) \mbox{d}t$
is the Debye function; see Abramowitz and Stegun~(1965; p.998).}
\label{tab:ACops}
\end{table}

\subsection{Constructing copulas by inversion (of Sklar's theorem)}
Beyond the bivariate case, copulas that are constructed through
inversion
of Sklar's theorem are popular; see Nelsen~(2006, Sect. 3.1). To derive
a copula function
in this way, let
$X=(X_1,\ldots,X_m) \in {\cal S}_X$ 
have 
distribution function
$G(x;\phi)$, with parameters $\phi$ and strictly monotonic
univariate marginal distribution functions
$G_1(x_1;\phi),\ldots,G_m(x_m;\phi)$. By Sklar's theorem, 
there always exists a copula
function $C$, such that
\[
G(x;\phi)=C(G_1(x_1;\phi),\ldots,G_m(x_m;\phi))\,.
\]
Denoting $u_j=G_j(x_j;\phi)$, then $x_j=G_j^{-1}(u_j;\phi)$, and
substituting this into the equation above defines a copula function:
\begin{equation}
C(u_1,\ldots,u_m;\phi)=
G(G_1^{-1}(u_1;\phi),\ldots,G_m^{-1}(u_m;\phi);\phi)\,.
\label{eq:ms:inv}
\end{equation}
It is important to notice that the multivariate distribution $G$ is only used 
to construct the copula function $C$, and
is not the distribution function of the random vector
$Y$, which remains $F$ as given in Equation~(\ref{eq:ms:sklar}).  
The parameters $\phi$ of the distribution of $X$ are the parameters
for copula function $C$.

Elliptical distributions are common choices for $G$ (Fang, Fang and Kotz~2002),
and the resulting copula functions are collectively called `elliptical copulas'.
The Gaussian
copula (Song~2000) is the most popular of these, where $G$ is the distribution function
of a multivariate normal with zero mean, correlation matrix
$\Gamma$ and unit variances
in each dimension. In this case, $\phi=\Gamma$, $G(x;\phi)=\Phi_m(x;\Gamma)$ and 
$G_j(x_j;\phi)=\Phi_1(x_j,1)$,
with $\Phi_k(\cdot;V)$ the distribution function of a $k$-dimensional $N(0,V)$ distribution.
The Gaussian copula function is therefore
\begin{equation}
C(u_1,\ldots,u_m;\phi)=\Phi_m(\Phi_1^{-1}(u_1;1),\ldots,\Phi_1^{-1}(u_m;1);\Gamma)\,.
\label{eq:ms:gcop}
\end{equation}
The restrictions on the first and second moments of $X$ are necessary to
identify the copula parameters $\Gamma$ in the likelihood.

When each marginal distribution $F_j$ is univariate normal
with mean $\mu_j$ and variance $\sigma_j^2$, then $u_j=\Phi_1(y_j-\mu_j;\sigma_j^2)$.
If a Gaussian copula is also assumed, then
the copula model for $Y$ simplifies to
a multivariate normal distribution
with mean $\mu=(\mu_1,\ldots,\mu_m)$ and covariance matrix $D\Gamma D$, with
$D=\mbox{diag}(\sigma_1,\ldots,\sigma_m)$.
 
Other choices for $G$ include a multivariate $t$ distribution, which 
results in the $t$ copula (Demarta and McNeil~2005), or a multivariate
skew $t$ distribution (Smith, Gan and Kohn~2010b). When selecting $G$, care
has to be taken to consider any 
restrictions on $\phi$ that
may be necessary to identify the parameters in the likelihood.

\subsection{Copula models as transformations}\label{sec:ms:cmtrans}
Copula modelling can be interpreted as a transformation from 
the domain of the data, to another domain where the dependence
is easier to model. The 
transformation is depicted in Figure~\ref{fig:trans}. If 
the elements of $Y$ are continuous-valued, the transformation $Y_j\mapsto U_j$
is one-to-one, as is the transformation $Y_j\mapsto X_j$ for inversion copulas.

The density of $Y$ is given by
\begin{equation}
f(y)=\frac{\partial}{\partial y} C(F_1(y_1),\ldots,F_m(y_m)) = c(u)\prod_{j=1}^m f_j(y_j)\,,
\label{eq:ms:cpdf}
\end{equation}
with $u=(u_1,\ldots,u_m)$, $u_j=F_j(y_j)$, $f_j(y_j)=\frac{\partial}{\partial y_j}F_j(y_j)$
and $c(u)=\frac{\partial}{\partial u}C(u)$.

However, when the data are discrete-valued, the probability mass function 
is obtained by differencing the distribution function in Equation~(\ref{eq:ms:sklar}), so that
\begin{equation}
\mbox{pr}(Y=y)=\Delta_{a_m}^{b_m}\Delta_{a_{m-1}}^{b_{m-1}}\cdots\Delta_{a_1}^{b_1}C(v)\,,
\label{eq:ms:pmf}
\end{equation}
where
$v=(v_1,\ldots,v_m)$ are indices of differencing.
The upper bound $b_j=F_j(y_j)$ and lower bound
$a_j=F_j(y_j^{-})$ is the left-hand limit of $F_j$ at $y_j$, with
$F_j(y_j^{-})=F_j(y_j-1)$ when $Y_j$ is ordinal-valued. In this
case
the transformations $Y_j \mapsto U_j$ and $Y_j \mapsto X_j$ are both
one-to-many. This means that the elements $U_j|Y_j=y_j$ and $X_j|Y_j=y_j$ are only 
known up to bounds, with
\begin{eqnarray*}
F_j(y_j^{-}) \leq & U_j & < F_j(y_j)\;\mbox{ and}, \\
G_j^{-1}(F_j(y_j^{-})) \leq & X_j & <G_j^{-1}(F_j(y_j))\,,
\end{eqnarray*}
for $j=1,\ldots,m$.
Nevertheless, $Y$, $U$ and $X$ still have distribution 
functions $F$, $C$ and $G$, respectively. 

It is
outlined later in Section~\ref{sec:ms:discrete}, how
interpreting a copula model as a
transformation allows for the construction of
Bayesian data augmentation schemes to evaluate the
posterior distribution when one or more margins are discrete.
\begin{figure}[tb]
\begin{center}
\begin{tabular}{lccccc}\hline
 & &$U_j=F_j(Y_j)$ & &$X_j=G_j^{-1}(U_j)$ & \\
Variable &$Y$ &$\longrightarrow$ &$U$ &$\longrightarrow$ &$X$ \\
Domain &$S_Y$ &$\longrightarrow$ &$[0,1]^m$ &$\longrightarrow$ &$S_X$ \\
Joint CDF &$F(y)$ &$\longrightarrow$ &$C(u)$ &$\longrightarrow$ &$G(x)$ \\
Marginal CDFs &$F_j(y_j)$ &$\longrightarrow$ &Uniform &$\longrightarrow$ &$G_j(x_j)$ \\
 \hline
\end{tabular}
\caption{Depiction of the transformation underlying a copula model.
The right hand
column for
variable $X$ is for copulas constructed by inversion only. The transformations
are given
in the top row for $Y_j$ continuous-valued.}
\label{fig:trans}
\end{center}
\end{figure}

\subsection{Vine copulas}\label{sec:ms:vines}
Much recent research in the copula literature has focused on 
building copulas in $m>2$ dimensions.
One popular
family of copulas are called `vines', which are constructed from sequences of bivariate
copulas. 
Joe~(1996; 1997) was an
early advocate of this approach, while Bedford and Cooke~(2002) organise the different
decompositions in a systematic way. Aas et al.~(2009) called the bivariate copulas `pair-copulas',
and vines are also known as pair-copula constructions (PCCs). 
Recent overviews are given by Haff, Aas and Frigessi~(2010)
and Czado~(2010).

Smith et al.~(2010) point out that if the elements of $Y$ are ordered in time, so that
$Y_t$ is observed before $Y_{t+1}$,
a vine labelled `decomposable' by Bedford and Cooke~(2002) (or D-vine for short)
proves a natural way of characterising serial dependence; particularly Markovian
serial dependence. This can be motivated by considering the 
following decomposition of the density of $U$,
\[
c(u)=\prod_{t=2}^m f(u_t|u_{t-1},\ldots,u_1)\,,
\]
where $f(u_1)=1$ because the marginal distribution of $u_1$ is uniform on $[0,1]$.
The idea is to build a representation for each conditional distribution
$f(u_t|u_{t-1},\ldots,u_1)$ as follows. For $s<t$ there always exists
a density $c_{t,s}$ on $[0,1]^2$ such that
\begin{eqnarray}
\lefteqn{f(u_t,u_s|u_{t-1},\ldots,u_{s+1})=f(u_t|u_{t-1},\ldots,u_{s+1})f(u_s|u_{t-1},\ldots,u_{s+1})} & &\nonumber \\
&\times &c_{t,s}\left(F(u_t|u_{t-1},\ldots,u_{s+1}),F(u_s|u_{t-1},\ldots,u_{s+1});
u_{t-1},\ldots,u_{s+1}\right)
\label{eq:ms:bivdecomp}
\end{eqnarray}
Here, $F(u_t|u_{t-1},\ldots,u_{s+1})$ and $F(u_s|u_{t-1},\ldots,u_{s+1})$ are
conditional distribution functions of $U_t$ and $U_s$, respectively. 
This is the theorem of Sklar applied conditional on $\{U_{t-1},\ldots,U_{s+1}\}$. 
In a vine copula, $c_{t,s}$ is the density of a bivariate `pair-copula' and it is
simplified by dropping dependence on $(u_{t-1},\ldots,u_{s+1})$;
see Haff, Aas and Frigessi~(2010) for a discussion of why this is often a good approximation. By setting 
$s=1$, application of Equation~(\ref{eq:ms:bivdecomp}) gives
\[
f(u_t|u_{t-1},\ldots,u_1)=c_{t,1}(F(u_t|u_{t-1},\ldots,u_2),F(u_1|u_{t-1},\ldots,u_2))f(u_t|u_{t-1},\ldots,u_2).
\]
Denoting $u_{t|j}=F(u_t|u_{t-1},\ldots,u_j)$ and $u_{j|t}=F(u_j|u_t,\ldots,u_{j+1})$, for $j<t$,
\footnote{Smith et al.~(2010) denote $u_{t|j}=F(y_t|y_{t-1},\ldots,y_j)$ and
$u_{j|t}=F(y_j|y_t,\ldots,y_{j+1})$ for $Y_1,\ldots,Y_m$ continuous random variables. However,
this can be shown to be equivalent to the definition of $u_{t|j}$ and $u_{j|t}$ employed here.}
repeated application of the above
with $s=2,3,\ldots,t-1$ leads to the following:
\[
f(u_t|u_{t-1},\ldots,u_1)=\prod_{s=1}^{t-1} c_{t,s}(u_{t|s+1},u_{s|t-1})\,,
\]
where the notation $u_{t|t}=u_t$, for $t=1,\ldots,m$. Therefore, the D-vine copula is given by
\begin{equation}
c(u)=\prod_{t=2}^m \left\{ \prod_{s=1}^{t-1} c_{t,s}(u_{t|s+1},u_{s|t-1}) \right\}\,,
\label{eq:ms:dvine}
\end{equation}
which is a product of $m(m-1)/2$ pair-copula densities, and $u=(u_{1|1},\ldots,u_{m|m})$.
If each pair-copula $c_{t,s}$ has copula parameter $\phi_{t,s}$, then the 
parameter vector of the D-vine is $\phi=\{\phi_{t,s}; t=2,\ldots,m,\,s<t\}$. 
The hardest aspect of using the copula in Equation~(\ref{eq:ms:dvine}) is the evaluation of the
arguments
of the component pair-copulas. Aas et al.~(2009),
give an $O(m^2)$ recursive algorithm for the evaluation of these from $u$,
based on the identity in Joe~(1996, p.125); see also Algorithm~1 in
Smith et al.~(2010).\footnote{The algorithm here corrects a minor subscript typographical
error in the algorithm in Smith et al.~(2010).}

\vspace{10pt}
\noindent \underline{Algorithm}: {\em (Evaluation of the Arguments of a D-vine)}
\vspace{5pt} \\
\noindent For $k=1,\ldots,m-1$ and $i=k+1,\ldots,m$:\\
\indent Step 1: Compute $u_{i|i-k}=h_{i,i-k}(u_{i|i-k+1}|u_{i-k|i-1};\phi_{i,i-k})$\\
\indent Step 2: Compute $u_{i-k|i}=h_{i,i-k}(u_{i-k|i-1}|u_{i|i-k+1};\phi_{i,i-k})$.
\vspace{10pt}

\noindent 
The functions $h_{t,s}(u_1|u_2;\phi_{t,s})=\int_0^{u_1} c_{t,s}(v,u_2;\phi_{t,s})\mbox{d}v$
are the conditional distribution functions for the pair-copula with density $c_{t,s}$;
see
Aas et al.~(2009) and Smith et al.~(2010) for lists of these for some
common bivariate copulas.

Because any combination of bivariate
copula functions can be employed for the pair-copulas, the D-vine copula can
be extremely flexible. Moreover, other
vine copulas can be constructed using alternative 
sequences of pair-copulas; see
Bedford and Cooke~(2002) and Aas et al.~(2009). However, 
the D-vine at Equation~(\ref{eq:ms:dvine}) 
is uniquely well-motivated
when the elements of $U$ are time-ordered.
 
\subsection{Measures of dependence}\label{sec:ms:mod}
Nelsen~(2006; Chap.5) and Joe~(1997; Chap.2) discuss measures of dependence
for copula models. In general, this is characterised by marginal pairwise dependencies
between elements $Y_i$ and $Y_j$. 
Kendall's tau and Spearman's rho are the two most popular
measures of pairwise concordance,
and empirical analysts are often familiar with sample versions 
based on ranked data. However, when $Y_i$ and $Y_j$ are
continuous-valued, and $Y$ follows
the copula model at Equation~(\ref{eq:ms:sklar}),
the population
equivalents can be expressed as
\begin{eqnarray}
\tau_{i,j} &= &4\left(\int_0^1\int_0^1 C^B_{i,j}(u_i,u_j)\mbox{d}C^B_{i,j}(u_i,u_j)\right)-1 =
4E(C^B_{i,j}(U_i,U_j))-1\,, \mbox{ and } \nonumber \\
\rho^S_{i,j} &= &12\int_0^1 \int_0^1 u_iu_j \mbox{d}C^B_{i,j}(u_i,u_j)-3 = 12E(U_iU_j)-3\,.
\label{eq:ms:depend}
\end{eqnarray}
In the above expressions, 
$C^B_{i,j}$ is the distribution function of $(U_i,U_j)$ and is
a bivariate margin of the $m$-dimensional copula
function $C$.
For some copulas $C^B_{i,j}$ can be computed
in closed form, but for others this is not possible. Similarly, the 
expectations in the expressions for $\tau_{i,j}$ and $\rho^S_{i,j}$
can sometimes be computed in closed form,
but for other choices of copulas they are computable only numerically, or
by Monte Carlo simulation. Within a Bayesian MCMC
framework the latter often proves straightforward; see Section~\ref{sec:ms:pinf}.

In many situations high values of $Y_i$ and $Y_j$ exhibit different levels 
(or even directions) of
dependence than low values of $Y_i$ and $Y_j$; something
that is called `asymmetric (pairwise) dependence'. 
As noted by Nelsen~(2006, Chap.4),
when $Y_i$ and $Y_j$ are continuous-valued, then the dependence properties
of the bivariate margin in these two variables is characterized by the
dependence properties between $U_i$ and $U_j$. 
In this case, measures of asymmetric
dependence are often based on the conditional probabilities
\begin{eqnarray*}
\lambda^{up}_{i,j}(\alpha) &= &\mbox{pr}(U_i>\alpha|U_j>\alpha) \\
\lambda^{low}_{i,j}(\alpha) &= &\mbox{pr}(U_i<\alpha|U_j<\alpha)\,,
\end{eqnarray*}
where $0<\alpha<1$.
The limits of these
are called the upper and lower tail dependencies (Joe~1997, p.33), and denoted as
\[
\lambda^{up}_{i,j}=\lim_{\alpha \uparrow 1}\lambda^{up}_{i,j}(\alpha)\,, \mbox{ and }
\lambda^{low}_{i,j}=\lim_{\alpha \downarrow 0}\lambda^{low}_{i,j}(\alpha)\,.
\]

For bivariate copula models there is only a single pairwise combination, $Y_1$ 
and $Y_2$,
and for many bivariate copula functions 
dependence measures are available
in closed form. For example, Table~\ref{tab:ACops} gives expressions for
measures of dependence for the Frank, Gumbel and Clayton copulas; see Joe~(1997),
Nelsen~(2006) and Huard, \'{E}vin and Favre~(2006) for others.
Pairwise dependence measures in multivariate $m$-dimensional elliptical copulas
can also have closed form expressions. In particular,
the Gaussian copula has zero tail dependence, with $\lambda^{up}_{i,j}=
\lambda^{low}_{i,j}=0$; whereas, the t copula
has tail dependence that is non-zero, but is symmetric with
$\lambda^{up}_{i,j}=\lambda^{low}_{i,j}$. When employing a copula model it is
important to ensure that the copula has dependence properties that are consistent with 
those exhibited by the data.

\section{Bayesian Inference for Continuous Margins}\label{sec:ms:bifcm}
When the data are continuous, the likelihood of $n$ independent observations
$y=\{y_1,\ldots,y_n\}$, each distributed as Equation~(\ref{eq:ms:sklar}),
is $f(y|\Theta,\phi)=\prod_{i=1}^n f(y_i|\Theta,\phi)$, where $y_i=(y_{i1},\ldots,y_{im})'$ and
\begin{equation}
f(y_i|\Theta,\phi)=c(u_i;\phi)\prod_{j=1}^m f_j(y_{ij};\theta_j)\,.
\label{eq:ms:like}
\end{equation}
Here, $u_i=(u_{i1},\ldots,u_{im})'$, $u_{ij}=F_j(y_{ij};\theta_j)$, $\Theta=\{\theta_1,\ldots,\theta_m\}$ are
any parameters of the marginal models, 
and $f_j(y_{ij};\theta_j)=\frac{\partial}{\partial y_{ij}} F_j(y_{ij};\theta_j)$ is the marginal 
density of $y_{ij}$. Initially, Equation~(\ref{eq:ms:like}) appears separable in $\theta_1,\ldots,\theta_m$
and $\phi$, but this
is not the case because $u_i$ depends on $\Theta$.
Most parametric copula functions have analytical expressions for the densities $c(u;\phi)$,
so that maximum likelihood estimation is often straightforward. However, there are 
a number of circumstances where
a Bayesian analysis can be preferable:
\begin{itemize}
\item[(i)] For more complex marginal models $F_j(y_{ij};\theta_j)$ and/or copula functions
$C(u;\phi)$,
the likelihood can be hard to maximise directly. One solution is to use a two
stage estimator, where the marginal model parameters $\theta_j$ are estimated first, and then
$\phi$ estimated conditional on these. 
In the copula literature, this is called `inference for margins'; see Joe~(2005) and
references therein for a discussion. Another solution is to
use to an iterative scoring algorithm to maximise the likelihood, as suggested
by Song, Fan and Kalbfleisch~(2006).
However, an
attractive Bayesian 
alternative in this circumstance is to construct inference
from the joint posterior $f(\Theta,\phi|y)$ evaluated in a Monte Carlo manner,
with $\Theta$ and $\phi$ generated separately in a Gibbs style sampling scheme; 
see Pitt, Chan and Kohn~(2006), Silva and Lopes~(2008) and Ausin and Lopes~(2010) for discussions.
\item[(ii)] Bayesian hierarchical modelling has proven very successful for the modelling
of multivariate data. This includes parsimonious modelling of
covariance structures using Bayesian selection and model averaging; see Giudici and Green~(1999),
Smith and Kohn~(2002), Wong, Carter and Kohn~(2003) and
Fr\"{u}hwirth-Schnatter \& T\"{u}chler~(2008) for
examples. Bayesian selection can be extended to nonlinear dependence by
considering priors with point mass components for $\phi$.
For example, 
Pitt, Chan and Kohn~(2006)
use a `spike and slab' prior similar to
Wong, Carter and Kohn~(2003) for the off-diagonal elements of the concentration matrix
$\Gamma^{-1}$ of a Gaussian copula.
Smith et al.~(2010) use Bayesian selection ideas to mix over
independent and
dependent pair-copulas in a vine copula. Hierarchical models can also be employed
for the margins $F_j(y_j;\theta_j)$, and estimated jointly with the dependence
structure captured by the copula function.
\item[(iii)] When estimating a copula model, the objective is often to construct inference on
measures of dependence, quantiles and/or
functionals of the random variable vector $Y$ or parameters
$(\Theta,\phi)$. Evaluation of the posterior distribution of 
these quantities is often straightforward using
MCMC methods. 
%For example, Smith et al.~(2010) outline how to compute the 
%Monte Carlo estimates of the marginal
%posterior distribution of
%Spearman's pairwise correlations $\rho^S_{ij}(\phi)$ for a vine copula.
\end{itemize}

\subsection{The Gaussian copula model}\label{sec:ms:Gcopmod}
To illustrate, Bayesian estimation of a Gaussian copula model for continuous
margins is outlined as
suggested by Pitt, Chan and Kohn~(2006).  
Following Song~(2000) and others, 
derivation of the copula density 
is straightforward by differentiation of Equation~(\ref{eq:ms:gcop}), so that
\begin{equation}
c(u;\phi)=\frac{\partial}{\partial u} C(u;\phi) =|\Gamma|^{-1/2}
\exp\left\{-\frac{1}{2} x'(\Gamma^{-1}-I)x \right\}\,,
\label{eq:ms:gcopdens}
\end{equation}
where $x=(\Phi^{-1}_1(u_1;1),\ldots,\Phi^{-1}_1(u_m;1))'$.
Thus, the likelihood at Equation~(\ref{eq:ms:like}) is a function of $\Theta$ and $\Gamma$, 
and can be written as
\begin{equation}
f(y|\Theta,\Gamma)=|\Gamma|^{-n/2}\left(\prod_{i=1}^n \exp
\left\{-\frac{1}{2} x_i'(\Gamma^{-1}-I)x_i\right\}
\prod_{j=1}^m f_j(y_{ij};\theta_j) \right)\,,
\label{eq:ms:gcoplike}
\end{equation}
where $x_i=(x_{i1},\ldots,x_{im})'$, $x_{ij}=\Phi^{-1}_1(u_{ij};1)$ 
and $u_{ij}=F_j(y_{ij};\theta_j)$.
Bayesian estimation can be undertaken using the following MCMC sampling scheme:

\vspace{10pt}
\noindent \underline{Sampling Scheme}: {\em (Estimation of a Gaussian Copula)}
\vspace{5pt} \\
\noindent Step 1: Generate from $f(\theta_j|\{\Theta\backslash \theta_j\},\Gamma,y)$ for $j=1,\ldots,m$.\\
\noindent Step 2: Generate from $f(\Gamma|\Theta,y)$.
\vspace{10pt}

Here, $\{A\backslash B\}$ is notation for $A$ with component $B$
omitted.
Steps~1 and~2 are repeated (in sequence) a large number of times, with each 
repeat
usually called a `sweep' in the Bayesian literature. The 
scheme requires an initial (feasible) state for the parameter values, which is denoted
here as $(\Theta^{[0]},\phi^{[0]})$. The iterates from the scheme form 
a Markov chain, which can be shown to converge to the joint posterior distribution
$f(\Theta,\phi|y)$, which is the (unique) invariant distribution of the chain. 
After an initial number of sweeps, the chain 
is assumed to have converged
and subsequent iterates form a Monte Carlo sample from which
the parameters are estimated, and other Bayesian inference obtained
as outlined in Section~\ref{sec:ms:pinf}. For introductions to MCMC
methods for computing Bayesian posterior inference see Tanner~(1996) and
Robert and Casella~(2006).

The posterior in Step~1 is given by
\begin{eqnarray}
\lefteqn{ f(\theta_j|\{\Theta\backslash \theta_j\},\Gamma,y) \propto f(y|\Theta,\Gamma)\pi(\theta_j) } \nonumber\\
 &\propto &|\Gamma|^{-n/2}\left(\prod_{i=1}^n \exp
\left\{-\frac{1}{2} x_i'(\Gamma^{-1}-I)x_i\right\} f_j(y_{ij};\theta_j)\right)
\pi(\theta_j)\,,
\label{eq:ms:margpost}
\end{eqnarray}
where $\pi(\theta_j)$ is the marginal prior for $\theta_j$. In general, the
density is unrecognisable
because $x_{ij}$ is a function of $\theta_j$, so
Pitt, Chan and Kohn~(2006) suggest using a Metropolis-Hastings (MH)
step with a multivariate $t$ distribution as a proposal to generate $\theta_j$ in Step~1.
The mean of the $t$ distribution, $\hat \theta_j$,
is the mode of Equation~(\ref{eq:ms:margpost}), which
is obtained via quasi-Newton-Raphson methods
applied to the logarithm of the posterior density.
The Hessian
\[
H=\left. \frac{\partial^2 \log(f(\theta_j|\{\Theta\backslash \theta_j\},\Gamma,y))}
{\partial \theta_j \partial \theta_j'} \right|_{\theta_j=\hat\theta_j}
\]
is calculated numerically
using finite difference methods. The scale matrix of the MH proposal is $-H^{-1}$, and a low
degrees of freedom, such as $\nu=5$ or $\nu=7$, is employed so that the proposal dominates
the target density in the tails. If $\theta_j$ has too many elements for 
$H$ to be evaluated in a numerically stable and computationally feasible fashion, 
$\theta_j$ can be partitioned and generated separately. Alternative MH 
steps are also possible, including those based on the widely employed
random walk proposals.

The approach used to generate $\Gamma$ in Step~2 varies depending on the prior and matrix
parameterisation adopted, of which there are several alternatives.
Pitt, Chan and Kohn~(2006) consider a prior on the off-diagonal elements of
$\Gamma^{-1}$, which is equivalent to assuming a 
prior for the partial correlations $\mbox{Corr}(X_t,X_s|X_{j \notin \{s,t\}})$
for $t=2,\ldots,m;\, s<t$.
Hoff~(2007) suggests using
a prior for $\Gamma$ in a Gaussian copula that results from an 
inverse Wishart prior for a covariance matrix.
However, because $\Gamma$ is just a correlation matrix (for $X$), any 
prior for a correlation matrix can also be used; 
for example, see those suggested by
Barnard, McCulloch and Meng~(2000), Liechty, Liechty and M\"{u}ller~(2004),
Armstrong et al.~(2009),
Daniels and Pourahmadi~(2009) and references therein.

\subsubsection{Prior based on a Choelsky factor:}
One such prior for a correlation matrix is based on a Cholesky factorisation, which is
particularly suited to longitudinal
data.
% see the discussion in Smith and Kohn~(2002).
This prior uses the decomposition 
\begin{equation}
\Gamma=\mbox{diag}(\Sigma)^{-1/2}\Sigma\, \mbox{diag}(\Sigma)^{-1/2}\,,
\label{eq:ms:cord}
\end{equation}
where $\Sigma$ is a non-unique positive definite matrix,
and $\mbox{diag}(\Sigma)$ is a diagonal
matrix comprised of the leading diagonal of $\Sigma$. The matrix
$\Sigma^{-1}=R'R$, with $R=\{r_{k,j}\}$ being an upper triangular Cholesky factor, and to ensure that the
parameterisation is unique, $r_{k,k}=1$, for $k=1,\ldots,m$.
Generation of $\Gamma$ in Step~2 is undertaken
by generating the elements $\{r_{k,j};j=2,\ldots,m,\, k<j\}$ one at a time from
the conditional posterior
\[ 
f(r_{k,j}|\{R\backslash r_{k,j}\},\Theta,y) \propto 
|\Gamma|^{-n/2}\left( \prod_{i=1}^n
\exp\left\{-\frac{1}{2} x_i'(\Gamma^{-1}-I)x_i\right\}\right)\pi(r_{k,j})\,,
\]
using random walk MH; see Tanner~(1996; p.177) for a discussion of this simulation tool. 
Once an iterate of $R$ is obtained, the iterate of $\Gamma$ can be computed using 
the relationship at Equation~(\ref{eq:ms:cord}). Using a different prior,
Hoff~(2007) uses a similar approach to generate a correlation matrix for a Gaussian copula. 

\subsubsection{Prior based on partial correlations:}
Daniels and Pourahmadi~(2009) suggest parameterising a correlation matrix using
the partial correlations
\begin{equation}
\lambda_{t,s}=\mbox{Corr}(X_t,X_s|X_{t-1},\ldots,X_{s+1})\,, \mbox{ for }s<t\,.
\label{eq:ms:pcors}
\end{equation}
This prior is based on the work of
Joe~(2006), who notes that these are unconstrained on $(-1,1)$,
and that $\Lambda=\{\lambda_{t,s}; t=2,\ldots,m,\,
s<t\}$ provides a unique parameterisation
of $\Gamma$. Note that $\lambda_{t,s}$ is sometimes called a `semi-partial' correlation
because it is not the correlation
conditional on all other variables $\mbox{Corr}(X_t,X_s|X_{j\notin \{t,s\}})$, which
is the `full' partial correlation considered by Pitt, Chan and Kohn~(2006). One advantage
is that the conditional distribution of $\lambda_{t,s}|\{\Lambda\backslash \lambda_{t,s}\}$ is only
bounded to $(-1,1)$, whereas
the conditional distribution of the full partial correlations
have more complex bounds.
Daniels and Pourahmadi~(2009) suggest using either Beta or uniform
priors for $\lambda_{t,s}$, which 
can be employed
and Step~2 undertaken
by generating the elements of $\Lambda$ one at a time, again using MH with
a random walk
proposal. Once an iterate of $\Lambda$ is obtained, $\Gamma$ can be computed using 
the identity at equation~(2) of Daniels and Pourahmadi~(2009).

There is an interesting link between the Gaussian copula parameterised by the partial
correlations $\Lambda$, and the 
D-vine copula in Equation~(\ref{eq:ms:dvine}).
When the pair-copulas in the D-vine are
bivariate Gaussian copulas, with densities
\begin{equation}
c_{t,s}(u_1,u_2; \phi_{t,s})\ = \frac{1}{\sqrt{1-\phi_{t,s}^2}}
\exp\left\{-\frac{\phi_{t,s}^2(x_1^2+x_2^2)-2\phi_{t,s}\,x_1\,x_2}{2(1-\phi_{t,s}^2)}\right\}\,,
\label{eq:ms:bivgauss}
\end{equation}
where $x_1=\Phi_1^{-1}(u_1;1)$ and $x_2=\Phi_1^{-1}(u_2;1)$, 
then the D-vine copula can be shown to be a
Gaussian copula with copula density at Equation~(\ref{eq:ms:gcopdens});
see Aas et al.~(2009) and
Haff, Aas and Frigessi~(2010). In this case, the individual pair-copula parameters $\phi_{t,s}$
above
are the partial correlations $\lambda_{t,s}$.

\subsection{Bayesian selection in a Gaussian copula}\label{sec:ms:bsgc}
Bayesian selection approaches can be employed
to allow for parsimonious modelling of $\Gamma$ in a Gaussian copula.
It is well known that Bayesian selection can significantly improve
estimates of a
covariance matrix compared to maximum likelihood; see
Yang and Berger~(1994),
Giudici and Green~(1998),
Smith and Kohn~(2002), Wong, Carter and Kohn~(2003),
Fr\"{u}hwirth-Schnatter \& T\"{u}chler~(2008)
and others
for extensive evidence to this effect. Pitt, Chan and Kohn~(2006) show that
this is also the case when estimating the dependence structure of $Y$ using a 
Gaussian copula model. They
consider a selection prior with point mass probabilities on the 
off-diagonal elements of $\Gamma^{-1}$.
In the Gaussian copula
this is equivalent to identifying for which pairs $(t,s)$ the full
partial
correlation $\mbox{Corr}(X_t,X_s|X_{j\notin \{s,t\}})=0$. 
This also corresponds to 
conditional independence between $Y_t$ and $Y_s$, 
with the conditional density 
$f(y_t,y_s|y_{j \notin \{s,t\}})=f(y_t|y_{j \notin \{s,t\}})f(y_s|y_{j \notin \{s,t\}})$.

\subsubsection{Priors for selection:}
Bayesian selection can also be undertaken for the semi-partial correlations $\Lambda$ defined
in Equation~(\ref{eq:ms:pcors}). In the Gaussian copula
this is equivalent to determining for which pairs $(t,s)$
there is conditional independence between elements of $Y$, with conditional density
\[f(y_t,y_s|y_{t-1},\ldots,y_{s+1})=
f(y_t|y_{t-1},\ldots,y_{s+1})f(y_s|y_{t-1},\ldots,y_{s+1})\,,
\]
when
$\lambda_{t,s}=0$. To introduce a point mass probability for this value,
binary
indicator variables $\gamma=\{\gamma_{t,s};t=2,\ldots,m,\, s<t\}$ are introduced,
such that
\[
\lambda_{t,s}=0 \mbox{ iff } \gamma_{t,s}=0\,.
\]
The non-zero partial
correlations $\lambda_{t,s}|\gamma_{t,s}=1$ are independently
distributed with proper prior densities $\pi(\lambda_{t,s})$.
Joe~(2006) highlights that
$\lambda_{t,s}|\{\Lambda \backslash \lambda_{t,s}\}$ are unconstrained
on $(-1,1)$, so that either independent uniform or Beta priors are simple
choices for $\pi(\lambda_{t,s})$; see Daniels and Pourahmadi~(2009).  In comparison, 
each
full partial correlation has bounds
that are complex functions of the other full partial correlations and computationally
demanding to evaluate.
For this reason, Bayesian
selection using the partial correlations $\Lambda$ is computationally less
burdensome than using the full partial correlations.

The prior on the indicators $\gamma$ can
be highly informative when the number of indicators $N=m(m-1)/2$ is large.
For example, if $w_\gamma=\sum_{t,s} \gamma_{t,s}$ is the number of non-zero
elements in $\Lambda$, then assuming flat marginal priors $\pi(\gamma_{t,s})=1/2$ 
puts high prior weight on values for $w_\gamma\approx N/2$. This problem
has been noted widely in the variable selection literature; see
Kohn, Smith and Chan~(2001), Zhang, Dai and Jordan~(2011)
and Bottolo and Richardson~(2010).
One solution is to employ the conditional prior 
\begin{equation}
\pi(\gamma_{t,s}=1|\{\gamma \backslash \gamma_{t,s}\}) \propto B(N-w_\gamma+1,w_\gamma+1)\,,
\label{eq:ms:cprior}
\end{equation}
where $B(\cdot,\cdot)$ is the beta function. This prior has been used effectively
in the Bayesian selection literature, with early uses in Smith~(2000)
and Smith and Kohn~(2002). It corresponds to assuming the
joint mass function
\[
\pi(\gamma)=\frac{1}{N+1}\left(\begin{array}{c} N \\ w_\gamma\end{array}\right)^{-1}\,.
\]
The implied
prior for the total number
of non-zero elements of $\Lambda$ is uniform, with $\pi(w_\gamma)=1/(1+N)$,
while
the
marginal priors $\pi(\gamma_{t,s})$ are all equal; see
Scott and Berger~(2010) for a discussion.
This prior is also equivalent to the uniform
volume-based prior suggested by Wong, Carter and Kohn~(2003) and 
Cripps, Carter and Kohn~(2005)
on the model space.

\subsubsection{MCMC sampling scheme:}
To evaluate the joint posterior distribution of the indicator variables
and the partial correlations $\Lambda$, 
%the sampling scheme   
%outlined in Smith et al.~(2010) for Bayesian 
%selection in a D-vine, can also be employed here. In this approach, 
latent variables
$\tilde \lambda_{t,s}$, for $t=2,\ldots,m, s<t$, are introduced such that 
$\lambda_{t,s}=\tilde \lambda_{t,s}$ if $\gamma_{t,s}=1$. Notice that $\lambda_{t,s}$
is known exactly given the pair $(\tilde \lambda_{t,s},\gamma_{t,s})$, so it is sufficient to
implement a sampling scheme to evaluate the joint posterior $f(\tilde \Lambda,\gamma,\Theta|y)$,
where $\tilde \Lambda=\{\tilde \lambda_{t,s};t=2,\ldots,m,\, s<t\}$, as below.

\vspace{10pt}
\noindent \underline{Sampling Scheme}: {\em (Bayesian Selection for a Gaussian Copula)} 
\vspace{5pt} \\
\noindent Step 1: Generate from $f(\theta_j|\{\Theta\backslash \theta_j\},\Gamma,y)$ for $j=1,\ldots,m$.\\
\noindent Step 2: Generate from $f(\tilde \lambda_{t,s},\gamma_{t,s}|\Theta,\{\tilde \Lambda
\backslash \tilde \lambda_{t,s}\},\{\gamma \backslash \gamma_{t,s}\},y)$ for $t=2,\ldots,m,\, s<t$.\\
\noindent Step 3: Compute $\Lambda$ from $(\tilde \Lambda, \gamma)$, and then $\Gamma$ from $\Lambda$.
\vspace{10pt}

Step~1 is unchanged from that in Section~\ref{sec:ms:Gcopmod}, while
Step~2
consists of MH steps to
generate each pair $(\tilde \lambda_{t,s},\gamma_{t,s})$, conditional on the others. The
MH proposal density is
\[
q(\tilde \lambda_{t,s},\gamma_{t,s})=q_1(\gamma_{t,s})q_2(\tilde \lambda_{t,s})\,.
\]
To generate from the proposal $q$ above,
an indicator is generated from $q_1(\gamma_{t,s}=0)=q_1(\gamma_{t,s}=1)=1/2$,
and
$\tilde \lambda_{t,s}$ from a symmetric 
random walk proposal $q_2$ constrained to $(-1,1)$. For example, one such symmetric proposal
for $q_2$ is to generate a new value of $\tilde \lambda_{t,s}$ from a normal distribution with mean
equal to the old value, standard deviation $0.01$, and constrained to $(-1,1)$.

Temporarily dropping the subscripts $(t,s)$ for convenience, a new iterate
$(\tilde \lambda^{new},\gamma^{new})$
generated from the proposal $q$ is
accepted over the old value
$(\tilde \lambda^{old},\gamma^{old})$
with probability
\begin{equation}
\min\left(1,\alpha 
\frac{\pi(\tilde \lambda^{new})}{\pi(\tilde \lambda^{old})}\kappa
\right)\,,
\label{eq:ms:mhrat}
\end{equation}
where $\kappa$ is an adjustment due to the bounds $(-1,1)$ on 
$\lambda$. If the symmetric density $q_2(\cdot)$
has distribution function $Q_2(\cdot)$, then 
\[
\kappa=\frac{Q_2(1-\tilde \lambda^{old})-Q_2(-1-\tilde \lambda^{old})} 
{Q_2(1-\tilde \lambda^{new})-Q_2(-1-\tilde \lambda^{new})} \,.
\]
If a uniform prior is adopted for $\tilde \lambda_{t,s}$,
as suggested in Daniels and Pourahmadi~(2009),
then
the ratio
$\pi(\tilde \lambda^{new})/\pi(\tilde \lambda^{old})=1$
in Equation~(\ref{eq:ms:mhrat}).
At each generation in Step~2, the likelihood
in Equation~(\ref{eq:ms:gcoplike}) is a function of
$(\tilde \lambda, \gamma)$, so it can be written here as $L(\tilde \lambda, \gamma)$.
Using this notation, the
value $\alpha$ in Equation~(\ref{eq:ms:mhrat})
can be expressed
separately for the four possible
configurations of $(\gamma^{old}, \gamma^{new})$
as:
\begin{eqnarray*}
\alpha\left((\tilde \lambda^{old},\gamma^{old}=0)\rightarrow (\tilde \lambda^{new},\gamma^{new}=0)\right)
&= &1\,, \\
\alpha\left((\tilde \lambda^{old},\gamma^{old}=0)\rightarrow (\tilde \lambda^{new},\gamma^{new}=1)\right)
&= &\frac{L(\tilde \lambda^{new},\gamma^{new}=1)\delta_1}{L(0,\gamma^{old}=0)\delta_0}\,, \\
\alpha\left((\tilde \lambda^{old},\gamma^{old}=1)\rightarrow (\tilde \lambda^{new},\gamma^{new}=0)\right)
&= &\frac{L(0,\gamma^{new}=0)\delta_0}{L(\tilde \lambda^{old},\gamma^{old}=1)\delta_1}\,,\\
\alpha\left((\tilde \lambda^{old},\gamma^{old}=1)\rightarrow (\tilde \lambda^{new},\gamma^{new}=1)\right)
&= &\frac{L(\tilde \lambda^{new},\gamma^{new}=1)}{L(\tilde \lambda^{old},\gamma^{old}=1)}\,,\\
\end{eqnarray*}
where $\delta_0$ and $\delta_1$ are the conditional probabilities from
Equation~(\ref{eq:ms:cprior}) that $\gamma_{t,s}=0$ and $1$, respectively. 
Notice that when 
$(\gamma^{old}=0) \rightarrow (\gamma^{new}=0)$ the likelihood does not need computing
to evaluate the acceptance ratio at Equation~(\ref{eq:ms:mhrat}).
This case will occur frequently whenever there is a high degree of sparsity in the dependence structure,
so that each sweep of Step~2 will be much faster than if no selection
was considered.

Reintroducing subscripts,
Step~3 of the sampling scheme is straightforward, with each partial correlation
\[
\lambda_{t,s}=\left\{ \begin{array}{ccc} 0 &\mbox{if} &\gamma_{t,s}=0\\ 
\tilde \lambda_{t,s} &\mbox{if} &\gamma_{t,s}=1 \end{array} \right. \,,
\]
and the
correlation matrix $\Gamma$ can be obtained directly from $\Lambda$ using the relationship
in Joe~(2006) and
Daniels and Pourahmadi~(2009).

\subsection{Bayesian estimation and selection for a D-vine}\label{sec:ms:bsdvine}
Bayesian estimation
for vine copulas is
discussed in Min and Czado~(2010; 2011) and
Smith et al.~(2010). The latter authors consider Bayesian selection and
model averaging
via the introduction of indicator variables in the tradition of Bayesian
variable selection. It is this
approach that is outlined here, although readers are referred to 
Smith et al.~(2010) for a full exposition.

The objective of Bayesian selection for a vine copula
is to identify component pair-copulas that are equal to the bivariate
independence copula. Recall that the bivariate independence copula
has copula function $C(u_1,u_2)=u_1u_2$, and corresponding copula density 
$c(u_1,u_2)=\partial C(u_1,u_2)/\partial u_1\partial u_2 =1$. This leads to a
parsimonious representation because the independence
copula is not a function of any parameters.

For
the D-vine with copula density at Equation~(\ref{eq:ms:dvine}),
Bayesian selection 
introduces indicator variables $\gamma=\{\gamma_{t,s}; t=2,\ldots,m,\,s<t\}$,
where
\begin{equation}
c_{t,s}(u_1,u_2)=\left\{ \begin{array}{cc}
1 &\mbox{ if }\gamma_{t,s}=0 \\
c_{t,s}^\star(u_1,u_2;\phi_{t,s}) &\mbox{ if }\gamma_{t,s}=1
\end{array} \right.\,.
\label{eq:ms:pcop}
\end{equation}
In the above, $c_{t,s}^\star$ is a pre-specified bivariate copula
density with parameter $\phi_{t,s}$.\footnote{Note that this parameter is
often a scalar, such as for an Archimedean or bivariate Gaussian copula. However,
it can also be a vector, as in the case of a bivariate t copula where 
both the degrees of freedom and correlation are parameters.}
The copula type can vary with
$(t,s)$, but for simplicity only the case where 
$c_{t,s}^\star(u_1,u_2;\phi_{t,s})=c^\star(u_1,u_2;\phi_{t,s})$ is considered here. 
That
is, each pair-copula $c_{t,s}$ is either an independence copula, or 
a bivariate copula of the same form for 
all pair-copulas, 
but with 
differing parameter values.
From Equation~(\ref{eq:ms:bivdecomp})
it follows that
when $c_{t,s}(u_1,u_2)=1$,  
$f(u_t,u_s|u_{t-1},\ldots,u_{s+1})=
f(u_t|u_{t-1},\ldots,u_{s+1})\times$ $f(u_s|u_{t-1},\ldots,u_{s+1})$, so that 
there is conditional independence between $U_t$ and $U_s$.

The pre-specified bivariate copula can
nest the independence copula, so that there exists a 
value $\phi^+$, such that  $c^\star(u_1,u_2;\phi^+)=1$.
In this
case,
the condition at Equation~(\ref{eq:ms:pcop}) can be rewritten as
$c_{t,s}(u_1,u_2)=c^\star(u_1,u_2;\phi_{t,s})$, with 
$\phi_{t,s} = \phi^+$ iff $\gamma_{t,s}=0$.
One example of such a copula is the Gumbel when
$\phi^+=1$, which is
easily seen by substituting the value into the copula
density, as given in Table~\ref{tab:ACops}.

To estimate the joint posterior
$f(\phi,\Theta|y)$,
latent variables
$\tilde \phi_{t,s}$, for $t=2,\ldots,m$, $s<t$, are introduced such that 
$\phi_{t,s}=\tilde \phi_{t,s}$ if $\gamma_{t,s}=1$. As with the partial correlations
in the previous section, $\phi_{t,s}$
is known exactly given the pair $(\tilde \phi_{t,s},\gamma_{t,s})$. Therefore,
it is sufficient to
implement a sampling scheme to evaluate the joint posterior 
$f(\tilde \phi,\gamma,\Theta|y)$,
where $\tilde \phi=\{\tilde \phi_{t,s};t=2,\ldots,m,\, s<t\}$, as below.

\vspace{10pt}
\noindent \underline{Sampling Scheme}: {\em (Bayesian Selection for a D-vine Copula)} 
\vspace{5pt} \\
\noindent Step 1: Generate from $f(\theta_j|\{\Theta\backslash \theta_j\},\phi,y)$ for $j=1,\ldots,m$.\\
\noindent Step 2: Generate from $f(\tilde \phi_{t,s},\gamma_{t,s}|\Theta,\{\tilde \phi
\backslash \tilde \phi_{t,s}\},\{\gamma \backslash \gamma_{t,s}\},y)$ for $t=2,\ldots,m,\, s<t$.\\
\noindent Step 3: Compute $\phi$ from $(\tilde \phi, \gamma)$.
\vspace{10pt}

Generating the marginal parameters $\theta_j$ in Step~1 
is undertaken using the same MH
step outlined in Section~\ref{sec:ms:Gcopmod}, 
but where
the conditional posterior is now
\begin{eqnarray*}
f(\theta_j|\{\Theta\backslash \theta_j\},\phi,y) &\propto & 
\left(\prod_{i=1}^n f(y_i|\Theta,\phi) \right)
\pi(\theta_j) \\
 &\propto &\left( \prod_{i=1}^n c(u_i;\phi)f_j(y_{ij};\theta_j) \right)
\pi(\theta_j)\,.
\end{eqnarray*}
In the above, $c(u_i;\phi)$ is the D-vine copula density at
Equation~(\ref{eq:ms:dvine}), evaluated at observation 
$u_i=(F_1(y_{i1};\theta_1),\ldots,F_m(y_{im};\theta_m))$.\footnote{In the copula
literature the $n$ observations $\{u_1,\ldots,u_n\}$ are often called the `copula data'.}
The algorithm in Section~\ref{sec:ms:vines} is run separately
for each observation $u_i$
to evaluate the arguments of the component pair-copulas of $c(u_i;\phi)$.
Interestingly, selection can speed up this algorithm substantially
because $h_{t,s}(u_1|u_2;\phi_{t,s})=u_1$ if $\gamma_{t,s}=0$.

Generating the pair ($\tilde \phi_{t,s},\gamma_{t,s})$ 
follows the same MH step outlined in Section~\ref{sec:ms:bsgc} for
the partial correlations. The main difference is that whenever
$\tilde \phi_{t,s}$ is vector-valued, each element is generated separately
in the same manner. Also, for many bivariate copulas (particularly
the Archimedean ones) proper non-uniform priors for 
$\tilde \phi_{t,s}$ 
are
often preferred.

\subsection{Equivalence of selection for Gaussian and D-vine copulas}
It is worth highlighting here that the Bayesian selection approach for the D-vine
nests that for the Gaussian copula, when the correlation matrix is parameterised by 
the semi-partial correlations $\Lambda$. 
If the pair-copula $c^\star$ is the bivariate Gaussian copula with
density at Equation~(\ref{eq:ms:bivgauss}), then $\phi_{t,s}=\lambda_{t,s}$  and
$\phi=\Lambda$. In this case, the sampling schemes for Bayesian selection for D-vine
and Gaussian copulas are identical.

\subsection{Posterior inference}\label{sec:ms:pinf}
Estimation is based on the Monte Carlo iterates
\[
\left\{(\phi^{[1]},\Theta^{[1]}),\ldots,
(\phi^{[J]},\Theta^{[J]})\right\}\,,
\]
obtained from the
sampling schemes after convergence
to the joint posterior distribution, so
that $(\phi^{[j]},\Theta^{[j]}) \sim f(\phi,\Theta|y)$. When Bayesian 
selection is undertaken, as in Sections~\ref{sec:ms:bsgc} and~\ref{sec:ms:bsdvine},
iterates $\{\gamma^{[1]},\ldots,\gamma^{[J]}\}$
are also obtained, with $\gamma^{[j]}\sim f(\gamma|y)$. Monte Carlo estimates
of the posterior means can be used as point estimates. For example,
the posterior means
\[
E(\theta_k|y) \approx \frac{1}{J}\sum_{j=1}^J \theta_k^{[j]}\,,\; \mbox{ and }
E(\phi|y) \approx \frac{1}{J}\sum_{j=1}^J \phi^{[j]}\,,
\]
are used as point estimates of the marginal model and copula parameters, respectively. 
Marginal $100(1-\alpha)\%$ 
posterior probability intervals can be constructed for any scalar parameter
by simply ranking the iterates, and then counting off the $\alpha J/2$ lowest
values, and the same number of the highest values.

When undertaking Bayesian selection for a Gaussian copula,
the estimates 
\[
\mbox{pr}(\gamma_{t,s}=1|y)\approx \frac{1}{J}\sum_{j=1}^J \gamma_{t,s}^{[j]}\,,
\mbox{ and }E(\lambda_{t,s}|y) \approx \frac{1}{J}\sum_{j=1}^J \lambda_{t,s}^{[j]}\,,
\]
can be computed. The former gives the posterior probability that the pair $Y_t,Y_s$
are independent, conditional on $(Y_{s+1},\ldots,Y_{t-1})$, for $s<t$. The
latter is the posterior mean of the semi-partial correlation.
At each
sweep of the sampling scheme, some elements of $\Lambda^{[j]}$ will be exactly equal to zero,
as determined by $\gamma^{[j]}$.
The estimate $E(\Gamma|y) \approx \frac{1}{J}\sum_{j=1}^J \Gamma^{[j]}$
is therefore often called a `model average'
because it 
is computed by averaging over these configurations of zero and non-zero semi-partial correlations
in $\Lambda^{[j]}$.

Similar estimates can be computed when undertaking Bayesian selection for D-vine copulas. 
When the form of the component pair-copulas nests the independence copula,
so that copula density $c^\star(u_1,u_2;\phi^+)=1$, then it is possible to compute
the posterior mean of the pair-copula parameters as
$E(\phi_{t,s}|y)\approx \frac{1}{J}\sum_{j=1}^J \phi_{t,s}^{[j]}$, because
$\phi_{t,s}^{[j]}=\phi^+$ when $\gamma_{t,s}^{[j]}=0$.
However, when the pair-copulas do not nest the independence copula,
$\phi_{t,s}$ is undefined when $\gamma_{t,s}=0$.

If the measures of pairwise dependence discussed in Section~\ref{sec:ms:mod} have a closed
form expression (or an accurate numerical approximation),
then Monte Carlo estimates are straightforward to compute. For example,
the estimate of Kendall's tau for continuous valued data is
\[
E(\tau_{i,k}|y) =\int \tau_{i,k}(\phi)f(\phi|y)\mbox{d}\phi \approx \frac{1}{J}\sum_{j=1}^J 
\tau_{i,k}(\phi^{[j]})\,.
\]
Posterior probability intervals are constructed
using the iterates $\{\tau_{i,k}(\phi^{[1]}),\ldots,$ $\tau_{i,k}(\phi^{[J]})\}$
in the same manner as for the model
parameters. If the pairwise dependence measures are difficult to compute, then
Kendall's tau and Spearman's rho
can be obtained by evaluating the expectations
at Equation~(\ref{eq:ms:depend}) via simulation as follows.
At the end of each sweep of a sampling scheme, generate an iterate
from the copula distribution $U^{[j]} \sim C(u;\phi^{[j]})$, and then
compute
\[E(C^B_{i,k}(U_i,U_k))\approx  \frac{1}{J}\sum_{j=1}^J C^B_{i,k}(U_i^{[j]},U_k^{[j]})\,,
\mbox{ and }
E(U_iU_k)\approx  \frac{1}{J}\sum_{j=1}^J U_i^{[j]}U_k^{[j]}\,.
\]
Simulating from most copula distributions is straightforward and fast; see
Cherubini, Luciano and Vecchiato~(2004; Chap.6).

\section{Bayesian Inference for Discrete Margins}\label{sec:ms:discrete}
Estimation of copula models with one or more discrete marginal distributions
differs substantially from those with continuous margins; see Genest and Ne\v{s}lehov\'{a}~(2007)
for an extensive discussion on the differences.
In this section,
the case where all margins are discrete is considered, although extension
to the case where some margins are discrete and others continuous
is discussed in 
Smith and Khaled~(2011).

The likelihood of $n$ independent observations
$y=\{y_1,\ldots,y_n\}$, each distributed as Equation~(\ref{eq:ms:sklar})
and with probability mass function at Equation~(\ref{eq:ms:pmf}), is
\begin{equation}
L(\Theta,\phi)=\prod_{i=1}^n \Delta_{a_{im}}^{b_{im}}\Delta_{a_{im-1}}^{b_{im-1}}\cdots\Delta_{a_{i1}}^{b_{i1}}C(v;\phi)\,.
\label{eq:ms:dlike}
\end{equation}
Here, $v=(v_1,\ldots,v_m)$ are indices of differencing,
each observation $y_i=(y_{i1},\ldots,y_{im})$, 
the upper bound $b_{ij}=F_j(y_{ij};\theta_j)$, and the lower bound
$a_{ij}=F_j(y_{ij}^{-};\theta_j)$ is the left-hand limit of $F_j$ at $y_{ij}$. 
In general, computing the likelihood involves $O(n2^m)$ evaluations of $C$,
which is prohibitive for high $m$.
Moreover, even for low values of $m$, it can be difficult to maximise
the likelihood for some copula and/or marginal model choices.

An alternative is to augment the 
likelihood with latent variables, and integrate them out
in a Monte Carlo fashion. From a Bayesian perspective this involves evaluating the 
augmented posterior distribution by MCMC methods; an
approach that is called Bayesian data augmentation
(Tanner and Wong~1987). Smith and Khaled~(2011)
discuss how this can be undertaken by augmenting the posterior distribution with
latent variables distributed as $U=(U_1,\ldots,U_m)\sim C(u;\phi)$.
While their approach 
applies to all parametric copula functions, 
in the specific case of a copula constructed by 
inversion as at Equation~(\ref{eq:ms:inv}), 
latent variables distributed as $X\sim G(x;\phi)$,
can also be used. Pitt, Chan and Kohn~(2006) propose
this to estimate Gaussian copula models, and
Smith, Gan and~Kohn~(2010b) when $G$ is the
distribution function of the skew t of
Sahu, Dey and Branco~(2003).

\subsection{The Gaussian copula model}
For the Gaussian copula, latent variables
$x=\{x_1,\ldots,x_n\}$ are introduced, where $x_i=(x_{i1},\ldots,x_{im})\sim N(0,\Gamma)$.
The augmented likelihood is $L(\Theta,\Gamma,x)=\prod_{i=1}^n f(y_i,x_i|\Theta,\Gamma)$, with
mixed joint density
\begin{eqnarray*}
f(y_i,x_i|\Theta,\Gamma) &= &\mbox{pr}(Y=y_i|x_i,\Theta)f_N(x_i;0,\Gamma)\\
 &= &\left(\prod_{j=1}^m I(A_{ij} \leq x_{ij} < B_{ij})\right)f_N(x_i;0,\Gamma)\,.
\end{eqnarray*}
Here, $f_N(x;\mu,V)$ is the density of a $N(\mu,V)$ distribution evaluated at $x$,
$I(Z)$ is an indicator function equal to one if $Z$ is true, and zero otherwise.
The mass function
\[
\mbox{pr}(Y_j=y_{ij}|x_{ij},\theta_j)=
\left\{\begin{array}{cc}1 &\mbox{ if }A_{ij}\leq x_{ij} < B_{ij} \\ 0 &\mbox{ otherwise}\end{array}\right.
\,,
\]
where
$A_{ij}=\Phi_1^{-1}(a_{ij};1)$ and $B_{ij}=\Phi_1^{-1}(b_{ij};1)$
as noted in
Section~\ref{sec:ms:cmtrans}, and $\Phi_1(\cdot;1)$ is the distribution function
of a standard normal. 

The likelihood of the copula model in Equation~(\ref{eq:ms:dlike})
is obtained by integrating over the latent variables, with
$L(\Theta,\Gamma)=\int L(\Theta,\Gamma,x) \mbox{d}x$.
Let $x_{(j)}=\{x_{1j},\ldots,x_{nj}\}$ be the latent variables corresponding to the
$j$th margin, then
the following sampling scheme can be used to evaluate the augmented posterior.

\vspace{10pt}
\noindent \underline{Sampling Scheme}: {\em (Data Augmentation for a Gaussian Copula)}
\vspace{5pt} \\
\noindent Step 1: For $j=1,\ldots,m$:\\
\indent 1(a) Generate from $f(\theta_j|\{\Theta\backslash \theta_j\},\{x\backslash x_{(j)}\},\Gamma,y)$\\
\indent 1(b) Generate from $f(x_{(j)}|\Theta,\{x\backslash x_{(j)}\},\Gamma,y)$\\
\noindent Step 2: Generate from $f(\Gamma|\Theta,x)$.
\vspace{10pt}

Steps 1(a) and 1(b) together produce an iterate
from the density\\
$f(\theta_j,x_{(j)}|\{\Theta\backslash \theta_j\},\{x\backslash x_{(j)}\},\Gamma,y)$.
The conditional posterior at Step~1(b) can be derived as
\begin{eqnarray*}
f(x_{(j)}|\Theta,\{x\backslash x_{(j)}\},\Gamma,y) &\propto &L(\Theta,\Gamma,x) \\ 
&\propto &\left( \prod_{i=1}^n I(A_{ij}\leq x_{ij} < B_{ij}) f_N(x_{ij};\mu_{ij},\sigma_{ij}^2)
\right)\,,
\end{eqnarray*}
where $\mu_{ij}$ and $\sigma_{ij}^2$ are the mean and variance of the conditional distribution
of $x_{ij}|\{x_i\backslash x_{ij}\}$ obtained from the joint distribution $x_i\sim N(0,\Gamma)$.
Thus, $x_{(j)}$ can be generated element-by-element from independent constrained normal densities.
In Step~1(a), $\theta_j$ is generated using the same MH approach
as in the continuous case, but where the conditional density is now
\[
f(\theta_j|\{\Theta\backslash \theta_j\},\{x\backslash x_{(j)}\},\Gamma,y)
\propto \left(\prod_{i=1}^n
\Phi_1\left(\frac{B_{ij}-\mu_{ij}}{\sigma_{ij}};1\right) -
\Phi_1\left(\frac{A_{ij}-\mu_{ij}}{\sigma_{ij}};1\right)\right) \pi(\theta_j)\,.
\]

In Step~2 any of the existing
methods for generating a correlation matrix $\Gamma$ from its posterior distribution
for Gaussian distributed data $x$ can be used, as 
outlined in Section~\ref{sec:ms:Gcopmod}. Bayesian selection ideas can also be used
as discussed in Section~\ref{sec:ms:bsgc}.

Pitt, Chan and Kohn~(2006) demonstrate the efficiency of this sampling scheme empirically, 
and Danaher and Smith~(2011)
show it can be applied effectively to a problem with $m=45$ dimensions.
Smith and Khaled~(2011) propose alternative sampling schemes that can be used
with the Gaussian copula, or
with other copula models.

\subsection{Measuring dependence}
For continuous multivariate data, dependence between elements of $Y$ is 
captured fully by the copula function $C$. In this case,
the measures of dependence based on $C$ discussed in Section~\ref{sec:ms:mod} are adequate
summaries. But when one or more margins are discrete-valued, in general,
measures of concordance involve the marginal distributions; see Denuit and Lambert~(2005),
and Ne\v{s}lehov\'{a}~(2007). Nevertheless, the
dependence structure of the latent vector $U$ (or the latent vector $X$ for copulas
constructed by inversion) is still informative concerning the level and type of
dependence in the data. Moreover, estimation using nonparametric rank-based estimators 
becomes inaccurate (Genest and Ne\v{s}lehov\'{a}~2007) and likelihood-based inference,
such as that outlined here, preferable.

\subsection{Link with multivariate probit and latent variable models}
Last, it is not widely appreciated that the multivariate probit model 
is a special case of the Gaussian copula model with univariate probit margins (Song~2000).
Data augmentation for a Gaussian copula therefore extends the approaches
of
Chib and Greenberg~(1998), Edwards and Allenby~(2003) and others for data augmentation
for a multivariate probit model,
to other Gaussian copula models. Similarly, the approach generalises a number of Gaussian
latent variable models for ordinal data, such as that of Chib and Winkelmann~(2001) and
Kottas, M\"{u}ller and Quintana~(2005).

\section{Discussion}
The impact of copula modelling in multivariate analysis has been substantial in many 
fields. Yet,
Bayesian inferential methods have been employed by only a few empirical analysts to date. Nevertheless,
they show great potential for computing efficient likelihood-based inference in a number of
of contexts. One of these
is in the modelling of multivariate discrete data, or data with a combination of discrete and 
continuous margins. Here, method of moments style 
estimators cannot be used effectively, and there can be computational difficulties in
maximising the likelihood,
so that Bayesian data augmentation becomes attractive; 
see Smith and Khaled~(2011)
for a full discussion.
Another is in the use of hierarchical models, including varying parameter models
(Ausin and Lopes~2010) or hierarchical models for Bayesian selection
and model averaging, as discussed here.
Last, while this article has focused on the Gaussian and
D-vine copulas, the Bayesian methods and ideas discussed here 
are applicable to a wide range of other
copula models, and it seems likely that their usage will increase in the near future.

\section*{Acknowledgements}
I would like to thank Robert Kohn, Claudia Czado, Anastasios Panagiotelis 
and particularly Mohamad Khaled, for their insightful comments on copula
models and associated methods of inference. This research was
funded by the Australian Research Council grants FT110100729 and
DP1094289.

\section*{References}
\parindent=0pt              
\begin{trivlist}
%\bibitem{aas09}
\item[]
Aas, K., Czado, C., Frigessi, A. and Bakken, H. (2009).
Pair-copula constructions of multiple dependence.
{\em Insurance: Mathematics and Economics}, {\bf 44}, 182--198.
%\bibitem{abramowitz}
\item[]
Abramowitz, M. and Stegun, I.A. (Eds.) (1965). 
{\em Handbook of Mathematical Functions},
New York: Dover Publications.

%\thebibliographybychapter{00}

\item[]%\bibitem{arm09}
Armstrong, H., Carter, C.K., Wong, K.F.K and Kohn, R. (2009).
Bayesian covariance matrix estimation using a mixture of 
decomposable graphical models.
{\em Statistics and Computing}, {\bf 19}, 303--316.

\item[]%\bibitem{ausin10}
Ausin, M.C. and Lopes, H.F. (2010). Time-varying joint distribution through copulas.
{\em Computational Statistics and Data Analysis}, {\bf 54}, 2383--2399.

\item[]%\bibitem{barnard00}
Barnard, J., McCulloch, R. and Meng, X. (2000). 
Modeling covariance matrices in terms of standard deviations and correlations, with 
application to shrinkage. {\em Statistica Sinica},
{\bf 10}, 1281--1311.

\item[]%\bibitem{bedford02}
Bedford, T. and Cooke, R. (2002). 
Vines - a new graphical model for dependent random variables.
{\em Annals of Statistics}, {\bf 30}, 1031--1068.

\item[]%\bibitem{bhat09}
Bhat, C.R. and Eluru, N. (2009). 
A Copula-Based Approach to Accomodation Residential Self-Selection Effects
in Travel Behavior Modeling.
{\em Transportation Research Part B}, {\bf 43}, 749--765.

\item[]%\bibitem{bottolo10}
Bottolo, L. and Richardson, S. (2010). 
Evolutionary Stochastic Search for Bayesian model exploration.
Preprint.

\item[]%\bibitem{cameron04}
Cameron, A., Tong, L., Trivedi, P. and Zimmer, D. (2004). 
Modelling the differences in counted outcomes using bivariate 
copula models with application to mismeasured counts.
{\em Econometrics Journal}, {\bf 7}, 566--584.

\item[]%\bibitem{chen06}
Chen, X. and Fan, Y. (2006). Estimation and model selection of semiparametric copula-based
multivariate dynamic models under copula misspecification.
{\em Journal of Econometrics}, {\bf 135}, 125--154.

\item[]%bibitem{chen06b}
Chen, X., Fan, Y. and Tsyrennikov, V. (2006). Efficient Estimation of Semiparametric Multivariate
Copula Models. 
{\em Journal of the American Statistical Association}, {\bf 101}, 1228--1240.

\item[]%\bibitem{cherubini04}
Cherubini, U. Luciano, E. and Vecchiato, W. (2004). 
{\em Copula methods in finance}, Wiley.

\item[]%\bibitem{chib98}
Chib, S. and Greenberg, E. (1998). Analysis of multivariate probit models.
{\em Biometrika}, {\bf 85}, 347--361.

\item[]%\bibitem{chib01}
Chib, S. and Winkelmann, R. (2001). Markov chain Monte Carlo Analysis of Correlated 
Count Data. {\em Journal of Business and Economic Statistics}, {\bf 19}, 428--435.

\item[]%\bibitem{clayton78}
Clayton, D. (1978). A model for association in bivariate
life tables and its application to epidemiological studies of family
tendency in chronic disease incidence. {\em Biometrika}, {\bf 65}, 141--151.

\item[]%\bibitem{cripps05}
Cripps, E., Carter, C., and Kohn, R. (2005). Variable Selection and Covariance
Selection inMultivariate Regression Models, in {\em Handbook of Statistics 25:
Bayesian Thinking: Modeling and Computation}, D. Dey and C. Rao, (Eds.),
Amsterdam: North-Holland, pp. 519--552.

\item[]%\bibitem{czado10}
Czado, C. (2010). Pair-Copula Constructions of Multivariate Copulas. In
{\em Workshop on Copula Theory and Its Applications}. Eds. F. Durante, W. H\"{a}rdle,
P. Jaworki, and T. Rychlik, Dordrecht: Springer.

\item[]%\bibitem{danaher11}
Danaher, P., and Smith, M. (2011). Modeling multivariate distributions
using copulas: applications in marketing, (with discussion).
{\em Marketing Science}, {\bf 30}, 4--21.

\item[]%\bibitem{daniels09}
Daniels, M. and Pourahmadi, M. (2009). Modeling covariance matrices via partial
autocorrelations. {\em Journal of Multivariate Analysis}, {\bf 100}, 2352--2363.

\item[]%\bibitem{demarta05}
Demarta, S. and McNeil, A.J. (2005).
The t-copula and related copulas. {\em International
Statistical Review}, {\bf 73}, 111--129.

\item[]%\bibitem{denuit05}
Denuit, M. and Lambert, P. (2005). Constraints on concordance measures
in bivariate discrete data. {\em Journal of Multivariate Analysis},
{\bf 93}, 40--57. 

\item[]%\bibitem{edwards03}
Edwards, Y.D. and Allenby, G.M. (2003). Multivariate Analysis of Multiple Response Data.
{\em Journal of Marketing Research}, {\bf 40}, 321--334.

\item[]%\bibitem{fang02}
Fang, H.B, Fang, K.T. and Kotz., S. (2002).
The Meta-elliptical Distributions with Given Marginals. {\em 
Journal of Multivariate Analysis}, {\bf 82}, 1--16.

\item[]%\bibitem{frees98}
Frees, E.W. and Valdez, E.A.  (1998).
Understanding Relationships Using Copulas. {\em North American Actuarial Journal}, {\bf 2}, 1--25.

\item[]%\bibitem{fruewirth08}
Fr\"{u}hwirth-Schnatter, S. and T\"{u}chler, R. (2008). 
Bayesian parsimonous covariance estimation for hierarchical linear mixed
models.
{\em Statistics and Computing}, {\bf 18}, 1--13.

\item[]%\bibitem{genest95}
Genest, C., Ghoudi, K. and Rivest, L.-P. (1995). A semiparametric estimation
procedure of dependence parameters in multivariate families of
distributions. {\em Biometrika}, {\bf 82}, 543--552.

\item[]%\bibitem{genest07}
Genest, C. and Ne\v{s}lehov\'{a}, J. (2007). 
A primer on copulas for count data.
{\em The Astin Bulletin}, {\bf 37}, 475--515. 

\item[]%\bibitem{genest93}
Genest, C. and Rivest, L.-P. (1993). Statistical Inference 
Procedures for Bivariate Archimedean Copulas.
{\em Journal of the American Statistical Association}, {\bf 88}, 1034--1043.

\item[]%\bibitem{guidici99}
Giudici, P. and Green, P. (1999).
Decomposable graphical Gaussian model determination.
{\em Biometrika}, {\bf 86}, 785--801.

\item[]%\bibitem{haff10}
Haff, I., K. Aas and Frigessi,A. (2010).
On the simplified pair-copula construction-
Simply useful or too simplistic?
{\em Journal of Multivariate Analysis},
{\bf 101}, 1296--1310. 

\item[]%\bibitem{hjort09} 
Hjort, N.L., Holmes, C., M\"{u}ller, P. and Walker, S. (2010).
{\em Bayesian Nonparametrics}, CUP.

\item[]%\bibitem{hoff07}
Hoff, P. (2007). Extending the rank likelihood for semiparametric copula estimation.
{\em The Annals of Applied Statistics}, {\bf 1}, 265--283.

\item[]%\bibitem{huard06}
Huard, D., \'{E}vin, G. and Favre, A.-C. (2006).
Bayesian copula selection.
{\em Computational Statistics and Data Analysis},
{\bf 51}, 809--822. 

\item[]%\bibitem{joe96}
Joe, H. (1996).
Families of m-variate distributions with given margins and m(m-1)/2 
bivariate dependence parameters. In R\"uschendorf, L., Schweizer, B.,
Taylor, M.D. (Eds.) {\em Distributions with Fixed Marginals and Related
Topics}.

\item[]%\bibitem{joe97}
Joe, H. (1997). {\em Multivariate Models and Dependence Concepts}, Chapman
and Hall.

\item[]%\bibitem{joe00}
Joe, H. (2005). Asymptotic efficiency of the two-stage estimation method
for copula-based models. {\em Journal of Multivariate Analysis}, {\bf 94}, 401--419.

\item[]%\bibtime{joe06}
Joe, H. (2006). Generating random correlation matrices based on partial correlations.
{\em Journal of Multivariate Analysis}, {\bf 97}, 2177--2189.

\item[]%\bibitem{kohn01}
Kohn, R., Smith, M. and Chan, D. (2001). Nonparametric regression using linear combinations of basis
functions. {\em Statistics and Computing}, {\bf 11}, 313-322.

\item[]%\bibitem{kottas05}
Kottas, A. M\"{u}ller, P. and Quintana, F. (2005). Nonparametric Bayesian Modeling for Multivariate
Ordinal Data. {\em Journal of Computational and Graphical Statistics}, {\bf 14}, 610--625.

\item[]%\bibitem{lambert02}
Lambert, P. and Vandenhende, F. (2002). A copula-based model for
multivariate non-normal longitudinal data: analysis of a dose titration safety
study on a new antidepressant. {\em Statistics in Medicine}, {\bf 21}, 
3197--3217.

\item[]%\bibitem{lee99}
Lee, A. (1999). Modelling Rugby League Data via Bivariate Negative Binomial Regression.
{\em Australian and New Zealand Journal of Statistics}, {\bf 41}, 141--152. 

\item[]%\bibitem{li00}
Li, D.X. (2000). On Default Correlation: A Copula Function Approach.
{\em The Journal of Fixed Income}, {\bf 9}, 43--54.

\item[]%\bibitem{liechty04}
Liechty, J.C., Liechty, M.W. and M\"{u}ller, P. (2004). 
Bayesian correlation estimation. {\em Biometrika},
{\bf 91}, 1--14.

\item[]%\bibitem{mcneil05}
McNeil, A.J., Frey, R. and Embrechts, R. (2005).
{\em Quantitative Risk Management: Concepts, Techniques and Tools},
Princeton University Press, Princton: NJ.

\item[]%\bibitem{min10}
Min, A. and Czado, C. (2010). Bayesian Inference for Multivariate
Copulas using Pair-Copula Constructions.
{\em Journal of Financial Econometrics}, {\bf 8}, 511--546.

\item[]%\bibitem{min11}
Min, A. and Czado, C. (2011). Bayesian model selection for D-vine pair-copula
constructions. 
{\em Canadian Journal of Statistics}, {\bf 39}, 239--258.

\item[]%\bibitem{nelsen06}
Nelsen, R.B. (2006). {\em An Introduction to Copulas}, (2nd Ed.), New York: Springer.

\item[]%\bibitem{neslehova07}
Ne\v{s}lehov\'{a}, J. (2007). On rank correlation measures for non-continuous
random variables. {\em Journal of Multivariate Analysis}, {\bf 98}, 544--567. 

\item[]%\bibitem{niko08}
Nikoloulopoulos, A. and Karlis, D. (2008). Multivariate logit copula model with
an application to dental data.
{\em Statistics in Medicine}, {\bf 27}, 6393--6406.

\item[]%\bibitem{niko10}
Nikoloulopoulos, A. and Karlis, D. (2010). Modeling multivariate count data using copulas.
{\em Communications in Statistics- Simulation and Computation}, {\bf 39}, 172--187.

\item[]%\bibitem{oakes89}
Oakes, D. (1989). Bivariate Survival Models Induced by Frailties.
{\em Journal of the American Statistical Association}, {\bf 84}, 487--493.

\item[]%\bibitem{oakes00}
Oakes, D. and Ritz, J. (2000). Regression in bivariate copula model.
{\em Biometrika}, {\bf 87}, 345--352.

%\item[]%\bibitem{panagiotelis08}
%Panagiotelis, A., and Smith, M. (2008). Bayesian Identification, Selection
%and Estimation of Semiparametric Functions in High-Dimensional Additive
%Models. {\em Journal of Econometrics}, {\bf 143}, 291--316.

\item[]%\bibitem{patton06}
Patton, A. (2006). Modelling Asymmetric Exchange Rate Dependence. {\em International
Economic Review}, {\bf 47}, 527--556. 

\item[]%\bibitem{pitt06}
Pitt, M., Chan, D. and Kohn, R. (2006).
Efficient Bayesian Inference for Gaussian Copula Regression Models.
{\em Biometrika}, {\bf 93}, 537--554.

\item[]%\bibitem{robert04}
Robert, C.P. and Casella, G. (2004). 
{\em Monte Carlo Statistical Methods}, 2nd Ed., New York: Springer 

\item[]%\bibitem{sahu03}
Sahu, S.K., Dey, D.K. and Branco, M.D. (2003).
A New Class of Multivariate Skew Distributions with Applications to Bayesian Regression Models.
{\em The Canadian Journal of Statistics}, {\bf 31}, 129--150.

\item[]%\bibitem{scott10}
Scott, J.G. and Berger, J.O. (2010). Bayes and empirical-Bayes multiplicity adjustment in the 
variable-selection problem. {\em The Annals of Statistics}, {\bf 38}, 2587--2619.

\item[]%\bibitem{shih95}
Shih, J.H. and Louis, T.A. (1995). 
Inferences on the Association Parameter in Copula Models for Bivariate Survival Data.
{\em Biometrics}, {\bf 51}, 1384--1399.

%\item[]%\bibitem{shively11}
%Shively, T.S., Walker, S.G. and Damien, P. (2011). 
%Nonparametric function estimation subject to monotonicity, convexity and other shape
%constraints.
%{\em Journal of Econometrics}, {\bf 161}, 166--181.

\item[]%\bibitem{silva08}
Silva, R. and Lopes, H., (2008). Copula, marginal distributions
and model selection: a Bayesian note.
{\em Statistics and Computing},
{\bf 18}, 313--320. 

\item[]
%\bibitem{sklar59}
Sklar, A. (1959). Fonctions de r\'epartition \`a n dimensions et
leurs marges.
{\em Publications de l'Institut de Statistique de L'Université de Paris},
{\bf 8}, 229--231.

\item[]
%\bibitem{smith03}
Smith, M.D. (2003). Modelling sample selection using Archimedean copulas.
{\em Econometrics Journal}, {\bf 6}, 99-123.

\item[]
%\bibitem{smith00}
Smith, M. (2000). Modeling and Short-Term Forecasting of New South Wales
Electricity System Load. {\em Journal of Business and Economic Statistics},
{\bf 18}, 465--478.

\item[]
%\bibitem{smithkauermann11}
Smith, M.S. and Kauermann, G. (2011). Bicycle commuting in Melbourne
during the 2000s energy crisis: A semiparametric analysis of intraday
volumes. {\em Transportation Research Part B}, forthcoming,
doi:10.1016/j.trb.2011.07.003.

\item[]
%\bibitem{smith11}
Smith, M.S. and Khaled, M.A. (2011). Estimation of Copula Models
with Discrete Margins via Bayesian Data Augmentation. 
{\em Journal of the American Statistical Association}, forthcoming.

\item[]
%\bibitem{smith02}
Smith, M. and Kohn, R. (2002). Parsimonious covariance matrix
estimation for longitudinal data, {\it Journal of the American Statistical
Association}, \textbf{97}, 1141--1153.

\item[]
%\bibitem{smith10}
Smith, M.S., Q. Gan and Kohn, R.J. (2010b). Modelling dependence using
skew t copulas: Bayesian inference and applications.
{\em Journal of Applied Econometrics}, (forthcoming), DOI: 10.1002/jae.1215.

\item[]
%\bibitem{smithmin10}
Smith, M., Min, A., Almeida, C., and Czado, C. (2010). Modeling longitudinal
data using a pair-copula decomposition of serial dependence.
{\em Journal of the American Statistical Association}, {\bf 105}, 1467--1479.

\item[]
%\bibitem{song00}
Song, P. (2000).
Multivariate Dispersion Models Generated from Gaussian
Copula. {\em Scandinavian Journal of Statistics}, {\bf 27}, 305--320.

\item[]
%\bibitem{song05}
Song, P., Fan, Y. and Kalbfleisch, J.D. (2005). Maximization by Parts in Likelihood
Inference. {\em Journal of the American Statistical Association}, {\bf 100}, 1145--1158.

\item[]
%\bibitem{tanner96}
Tanner, M.A. (1996). {\em Tools for Statistical Inference: Methods for the Exploration of Posterior
Distributions and Likelihood Functions}, 3rd Ed., New York: Springer.

\item[]
%\bibitem{tanner87}
Tanner, M.A. and Wong, W.H. (1987). The Calculation of Posterior Distributions by Data
Augmentation. {\em Journal of the American Statistical Association}, {\bf 82}, 528--540.

\item[]
%\bibitem{wong03}
Wong, F., Carter, C.K. and Kohn, R. (2003). Efficient estimation of covariance selection models. 
{\em Biometrika}, {\bf 90}, 809--830.

\item[]
%\bibitem{yang94}
Yang, R. and Berger, J.O. (1994). Estimation of a Covariance Matrix Using the Reference Prior.
{\em Annals of Statistics}, {\bf 22}, 1195--1211.

\item[]
%\bibitem{zellner62}
Zellner, A. (1962). An Efficient Method of Estimating Seemingly Unrelated Regressions
and Tests for Aggregation Bias.
{\em Journal of the American Statistical Association}, {\bf 57}, 348--368.

\item[]
%\bibitem{zhang11}
Zhang, Z., Dai, G. and Jordan, M. (2011). Bayesian Generalized Kernel Mixed Models.
{\em Journal of Machine Learning Research}, {\bf 12}, 111--139.
\end{trivlist}

%\endthebibliographybychapter

\end{document}